\documentclass[traditabstract]{aa} 

\usepackage{epsfig}
\usepackage{graphicx}
\usepackage{txfonts}
\usepackage{amssymb}
\usepackage{natbib}                               
\def\apjl{ApJL }

\def\apjs{ApJS }

\def\aap{A\&A }

\def\I{{\em INTEGRAL}} 
\def\rxte{{\em RXTE}}

\def\be{\begin{equation}} 
\def\ee{\end{equation}} 
 
\begin{document} 
\title{The ephemeris, orbital decay, and masses of 10 eclipsing HMXBs} 
 
\author{M. Falanga\inst{1,2}
\and E. Bozzo\inst{3}
\and  A. Lutovinov\inst{4}
\and J. M. Bonnet-Bidaud\inst{5}
\and Y. Fetisova\inst{4}
\and J. Puls\inst{6}
} 
 
\titlerunning{The ephemeris, orbital decay, and masses of 10 HMXBs}  
\authorrunning{Falanga et al.}  
  
\institute{International Space Science Institute, Hallerstrasse 6, 3012 Bern, Switzerland 
 \email{mfalanga@issibern.ch}
\and International Space Science Institute Beijing, No.1 Nanertiao, Zhongguancun, Haidian District, 100190 Beijing, China
\and ISDC Data Centre for Astrophysics, Chemin d'\'{E}cogia 16, 1290, Versoix, Switzerland
\and Space Research Institute, Russian Academy of Sciences, Profsoyuznaya 84/32, 117997 Moscow, Russia
\and Service dAstrophysique (SAp), IRFU/DSM/CEA-Saclay, 91191 Gif-sur-Yvette Cedex, France
\and Universit\"atssternwarte der Ludwig-Maximilians-Universit\"at M\"unchen, Scheinerstrasse 1, 81679 M\"unchen, Germany
} 
 
\date{ } 
 
  \abstract 
{We take advantage of more than 10 years of monitoring of the eclipsing HMXB systems LMC X-4, 
Cen X-3, 4U 1700-377, 4U 1538-522, SMC X-1, IGR J18027-2016, Vela X-1,IGR J17252-3616, XTE J1855-026, and OAO 1657-415 with 
the ASM on-board RXTE and ISGRI on-board INTEGRAL to update their ephemeris. These results are used to refine 
previous measurements of the orbital period decay of all sources (where available) and provide the first accurate 
values of the apsidal advance in Vela\,X-1 and 4U\,1538-522. Updated values for the masses of the neutron 
stars hosted in the ten HMXBs are also provided, as well as the long-term lightcurves folded on the sources best determined orbital 
parameters. These lightcurves reveal complex eclipse ingresses and egresses, that are understood mostly as being due to the presence of accretion 
wakes. The results reported in this paper constitute a database to be used for population and evolutionary studies of 
HMXBs, as well as theoretical modelling of long-term accretion in wind-fed X-ray binaries.}  

\maketitle

\section{Introduction} 
\label{sec:intro} 

High-mass X-ray binaries (HMXB) are among the brightest X-ray sources in our Galaxy, and were discovered for the first time  
with the {\em Uhuru} satellite \citep{Giacconi1971} and independently with balloon-born instruments \citep[see e.g.,][]{Lewin1971}. 
These sources typically host a compact object (often a neutron star, NS) accreting material lost by a massive 
companion. Depending on the nature of the latter star, HMXBs are divided into Be or Supergiant X-ray binaries (SgXBs). In the first case, 
a conspicuous X-ray emission is generated due to the accretion of the matter from the dense decretion disk around the companion 
star, that leads to transient events (outbursts) of different types \citep[see e.g.,][for a review]{Reig2011}. 
In the SgXBs, the compact object can either accrete through the fast and dense wind produced by the companion, 
or in a few cases through the so-called Roche-lobe overflow \citep[a number of systems showed evidence for both mechanisms 
contributing to the accretion onto the compact object and the overall X-ray emission; see, e.g.,][for a recent review]{chaty11}.   
\begin{table*}[ht!]
\caption{\label{table1:sources}Orbital parameters of the 10 eclipsing HMXB systems in ascending order of the orbital 
period. Where relevant, we indicated in brackets the uncertainties on the last digits of each reported value.}
\begin{center}
\begin{tabular}{lcccccccc}
\hline 
Source & Type & Orbital Epoch$^{\rm m}$   & $P_{\rm orb}$ & $P_{\rm s}$ & $\dot P_{\rm orb}/P_{\rm orb}$ &  $a_{\rm x}$ sin $i$ & $e$& $\omega$\\
 & & (MJD) & (days)              & (sec)       &  $10^{-6}$ yr$^{-1}$  & (lt-sec)                  & &(deg)\\
\hline 
\hline 
LMC X-4$^{\rm a}$ & O8III & 51110.86579(10) &  1.40839776(26) &  13.5 & -0.98(7)  &  26.343(16)&$0.006(2)$ &--\\
Cen X-3$^{\rm b}$ & O(6-7)II-III & 50506.788423(7) &  2.087113936(7)  & 4.8  & -1.799(2) & 39.6612(9) &$<0.0016$ &--\\
4U 1700-377$^{\rm c}$ & O6.5Iaf+ & 48900.373(15) & 3.411581(27) &  -- & -3.3(6) & 48--82  &--$^{\rm l}$ &49(11)\\
4U 1538-522$^{\rm d}$ & B0.2Ia & 52851.33(1) & 3.728382(11) & 526.8 & -- & 53.1(1.5) & 0.18(1) &40(12)\\
SMC X-1$^{\rm e}$ & B0sg & 50324.691861(8) & 3.89220909(4) & 0.71  & -3.402(7) & 53.4876(9) & 0.00089(6) & 166(12)\\
SAX J1802.7-2017$^{\rm f}$ & B1Ib & 52168.26(4) & 4.5696(9) &  139.6 & -- & 68(1)& -- &--\\
XTE J1855-026$^{\rm g}$ & B0Iaep & 51495.25(2) & 6.0724(9) & 360.7 & -- & 80.5(1.4) & 0.04(2) & 226(15)\\
Vela X-1$^{\rm b}$ & B0.5Iae & 48895.2186(12) & 8.964368(40) & 283.2  & -- & 113.89(13) &  0.0898(12) &152.59(92)\\
EXO 1722-363$^{\rm h}$ & B0-B1Ia & 53761.68(4) & 9.7403(4) & 413.9 & --& 101(1) & $<0.19$ &--\\
OAO 1657-415$^{\rm i}$ & B0-6sg & 50689.116(50) & 10.44749(55) & 37.3 & -3.40(15) & 106.10(2) & 0.1033(6) & 87.6(1.3)\\
\hline
\end{tabular}  
\end{center}
\tablefoot{$^{\rm a}$\citet[][]{Levine2000,Bildsten1997};  $^{\rm b}$\citet{Bildsten1997,Raichur2010};  
$^{\rm c}$\citet{Rubin1996,Hammerschlag-Hensberge2003}; 
$^{\rm d}$\citet{Mukherjee2006,Clark2000}; $^{\rm e}$\citet{Inam2010,Raichur2010}; $^{\rm f}$(a.k.a. IGR J18027-2016) \citet{Hill2005}; 
$^{\rm g}$\citet{Corbet2002}; $^{\rm h}$(a.k.a. IGR J17252-3616) \citet{ZuritaHeras2006,Thompson2007,Manousakis2011}; 
$^{\rm i}$\citet{Baykal2011,Jenke2011}. $^{\rm l}$The eccentricity for this source has been reported by \citet{Hammerschlag-Hensberge2003} 
to be 0.22(4) and later questioned by \citet{Clark2000} and this work. $^{\rm m}$Mid-eclipse time, equivalent to time 
when mean longitude $l$ equals $\pi/2$ for a circular orbit.}
\end{table*}

In Roche-lobe accreting systems, the inflowing material from the companion star is endowed with a high specific angular momentum, and thus 
accretion is generally mediated by a disk surrounding the compact object. As this is a very efficient 
way to accrete matter, disk-fed systems achieve usually a relatively high X-ray luminosity ($\sim 10^{38}$ erg s$^{-1}$). 
In wind accreting systems, the compact object is instead deeply embedded in 
the material lost by the companion star, which do not posses sufficient angular momentum to form an accretion disk 
and accretes quasi-spherically onto the NS \citep[see, e.g.][for recent reviews]{Bozzo2008,Shakura2012}. This process lead to 
a strongly variable X-ray emission, reaching values significantly lower than those attained by disk-accreting system 
\citep[ranging from $10^{34}$ to $10^{36}$~erg~s$^{-1}$; see, e.g.,][]{Joss1984,Nagase1996,Bildsten1997}.

Since their discovery, HMXBs have been intensively monitored in order to test wind accretion models onto 
magnetized NSs \citep[see e.g.,][and references therein]{Lamers1976,Bhattacharya1991,Lutovinov2013}, measure the long term spin-variation of 
NSs as function of the mass accretion rate \citep[probing the so-called ``accretion-torques'', see e.g.,][]{Ghosh1979,Lovelace1995}, 
and study the evolution of the orbital parameters of these system through, e.g. measurements of the orbital period decay 
\citep[see][and references therein]{Bildsten1997}. In the majority of HMXBs, pulsations in the X-ray emission firmly established the presence 
of NSs as accreting compact objects \citep[see e.g.,][for a recent review and HMXBs statistic]{Lutovinov2005,Lutovinov2009}, 
and the detection of cyclotron absorption spectral features permitted the direct measurement 
of their magnetic field strength \citep[$\sim1-5\times10^{12}$~G; see, e.g.,][]{Coburn2002,Filippova2005,Caballero2012}. 
In the HMXBs that did not show clear evidence of pulsations, the presence of black holes companions cannot still be firmly ruled 
out \citep[see, e.g., the case of 4U\,1700-377,][]{Rubin1996,Clark2002}. 

Among more than hundred HMXBs, only  in a few sources  
 the inclination of the system is high enough that the compact star is periodically occulted along our line of sight 
by the companion, giving rise to X-ray eclipses. For these sources, it is possible to infer a number of orbital 
parameters (e.g., orbital period) from the measured duration of the eclipse. The energy-dependent profile of the X-ray lightcurve during the 
eclipse ingress and egress also reveal details of the OB stellar wind structure \citep[e.g.,][]{White1995}. 
In those systems in which orbital ephemerides can be accurately measured over several decades, it is expected that changes in the orbital 
period can also be revealed, providing further insight into tidal interaction and mass transfer mechanisms that regulated the accretion 
onto the compact object. In eccentric systems, accurate ephemerides can also be used to infer the angle of the periastron 
and the apsidal advance \citep[see e.g.,][and references therein]{vanderKlis1984}. 

In this work we take advantage of the available long-term monitoring observations of ten 
eclipsing HMXBs\footnote{Be X-ray binaries are not expected to show long detectable X-ray eclipses, as the radial extent of 
the companion is less than those of supergiant stars in HMXBs and the orbital periods are usually much longer and 
characterized by high eccentricities \citep{Reig2011}.}, carried out with the hard X-ray imager IBIS/ISGRI (17~keV -- 200~keV) 
on-board {\em INTEGRAL} and the All Sky Monitor (ASM) on-board {\em RXTE} (2--12~keV), to obtain the most accurate 
ephemeris available so far for these sources \citep{Liu2006}. We significantly improved their previously measured orbital periods, 
eclipse durations, and masses of the accreting NSs hosted in them.  Measurements of the orbital period decay and apsidal 
motion is presented for the first time for several of these systems, and refined measurements are provided for all the others with 
respect to those already available in the literature. We discuss the energy-dependent  
profile of the long-term lightcurve during the eclipse ingress and egress for all these systems in terms of X-ray absorption  
expected in wind and/or disk accreting systems.

\subsection{The 10 eclipsing HMXBs}
\label{sec:sources} 

In Table \ref{table1:sources} we list the ten sources selected for the present work and provide a summary of their main characteristics, 
as measured in the most recently available literature: the orbital epoch, the orbital period $P_{\rm orb}$, the spin period $P_{\rm s}$, 
the orbital period change  $\dot P_{\rm orb}/P_{\rm orb}$, the projected semi-major axis $a_{\rm x}\sin\, i$, the eccentricity $e$, 
and the periastron angle $\omega$. All these systems show evidence for alternate spin-up and spin-down phases. In some cases 
the average spin evolution led to an effective long term increase or decrease of the NS spin rate 
\citep[see e.g.,][for more details]{Bildsten1997,Inam2010}, and thus the spin periods reported in Table~\ref{table1:sources} 
should only be considered as indicative. The spin period of the NS (possibly) hosted in 4U1700-377 is currently unknown. 

For the two HMXBs EXO 1722-363 and SAX J1802.7-2017, we indicated in Table~\ref{table1:sources} also the associated 
{\em INTEGRAL} source name IGR~J17252-3616 and IGR~J18027-2016, respectively \citep{ZuritaHeras2006, Revnivtsev2004}. 
These two sources are classified as highly absorbed HMXBs \citep[see e.g.,][]{Walter2006}, and indeed these are the only sources 
in which no eclipse is evident from their soft X-ray lightcurves ($<$12 keV, see Sec.~\ref{sec:folded_lc}). 

The 10 selected sources comprise both disk and wind accreting systems. A different shape of the eclipse ingress and egress 
is expected in these two cases as a function of the energy (see Sec.~\ref{sec:folded_lc}). The  relatively large 
X-ray luminosity of LMC X--4, Cen X--3, and SMC X--1  makes  these sources the prime candidates for being disk-fed systems 
whereas 4U 1700--377, 4U 1538--522, SAX J1802.7--2017, 
XTE J1855--026, Vela X--1, and EXO 1722--363 are all thought to be wind-fed systems. 
OAO 1657--415 is unique among the known HMXBs as it is believed to be a wind-fed system for most of the time, and undergo 
only sporadically episodes of accretion from a temporary disk. Since the binary is too wide for Roche lobe overflow to occur, 
this may provide the first clear evidence that winds in HMXBs possess sufficient angular momentum to form accretion disks 
\citep[see e.g.,][]{Bildsten1997,Chakrabarty1993}. Further additional information on the ten sources can be found 
in \citet{Liu2006}.

\section{Observations and data} 
\label{sec:data}   

All the eclipsing HMXBs reported in Table \ref{table1:sources} have been continuously  
monitored in the X-ray domain by the ASM since the beginning of 1996 
and by the IBIS/ISGRI since the early 2003. 
We used public available IBIS/ISGRI lightcurves retrieved from the on-line tool High-Energy Astrophysics 
Virtually ENlightened Sky ({\it HEAVENS})\footnote{http://www.isdc.unige.ch/heavens/}. 
{\it HEAVENS} data reduction was performed using the standard Offline Science Analysis 
(OSA) version 9.0 distributed by the {\it INTEGRAL} Science Data Center \citep{Courvoisier2003}. 
For each source we downloaded the ISGRI high energy lightcurve binned over each pointing 
(science window) of roughly 2 ks, in the 17--40 keV and 40--150 keV band. 
The \rxte/ASM lightcurves have been retrieved from the NASA HEARSAC FTP 
server\footnote{ftp://legacy.gsfc.nasa.gov/xte/data/archive/ASMP} binned dwell-by-dwell 
(90 s bins) in the 1.5--3 keV, 3--5 keV, and 5--12 keV energy bands. 
In Table \ref{table2:log}, we report the observation time interval and the total 
effective exposure time of the ASM and ISGRI data for each of the selected source. 
All the photons arrival times of the \I/ISGRI and \rxte/ASM lightcurves were corrected 
to the barycenter reference time of the solar system. We used the OSA task {\tt barycent} for the 
ISGRI data, and the procedure described in the \rxte\ 
cookbook\footnote{http://heasarc.nasa.gov/docs/xte/recipes/asm\_recipe.html\#barycenter} for the ASM data. 
The barycentric correction is usually considered mainly to perform highly precision 
timing analysis on a short observational time interval. However, in the present case, 
it was required in order to ensure uniformity over a long-term set of data spanning 
more than a decade. 

For each of the eclipse found in the ISGRI and ASM lightcurves we measured the ingress, egress and 
the mid-eclipse time as described in Sec.~\ref{sec:orbital_priod}.
\begin{table}[hbt]
\tiny
%\begin{center}
\caption{\label{table2:log} Log of \I/ISGRI and \rxte/ASM observations of the 10 selected eclipsing HMXBs.}
\begin{tabular}{llrll}
\hline 
 Source &ISGRI    & Exp.  & ASM & Exp.\\
 & (MJD-50000)& (ks) & (MJD-50000) & (ks)\\
\hline 
\hline 
LMC X-4 & 2641.40--5005.48 &  659 & 87.30--5748.11 & 5627\\
Cen X-3 & 2650.54--5155.90 & 2033& 87.29--5747.29 & 5297\\
4U 1700-377 & 2668.25--5256.35 & 2863 & 88.11--5749.81& 4709\\
4U 1538-522 &  2650.74--5256.37 & 2285 & 88.11--5749.69 & 5494\\
SMC X-1 & 2843.66--5008.01 & 704 & 88.35--5747.51 & 5405\\
SAX J1802.7-2017 & 2698.16--5256.18 & 4037 & 94.24--5744.55 & 2707\\
XTE J1855-026 & 2704.14--5136.03 & 2482 & 88.37--5749.75 & 5409\\
Vela X-1 & 2644.45--5150.95 & 1665 & 87.29--5747.53 & 6317\\
EXO 1722-363 & 2668.25--5256.35 & 4147 & 88.11--5749.81 & 4657\\
OAO 1657-415 & 2668.25--5256.37 & 2353 & 91.13--5747.53 & 5065\\
\hline
\end{tabular}
%\end{center}
\end{table}

\section{Best fit ephemerides}
\label{sec:ephemerides}

\subsection{Orbital period and orbital period decay}
\label{sec:orbital_priod}

\begin{figure*}
  \begin{center}
   \begin{tabular}{ccc}
      \resizebox{55mm}{!}{\includegraphics{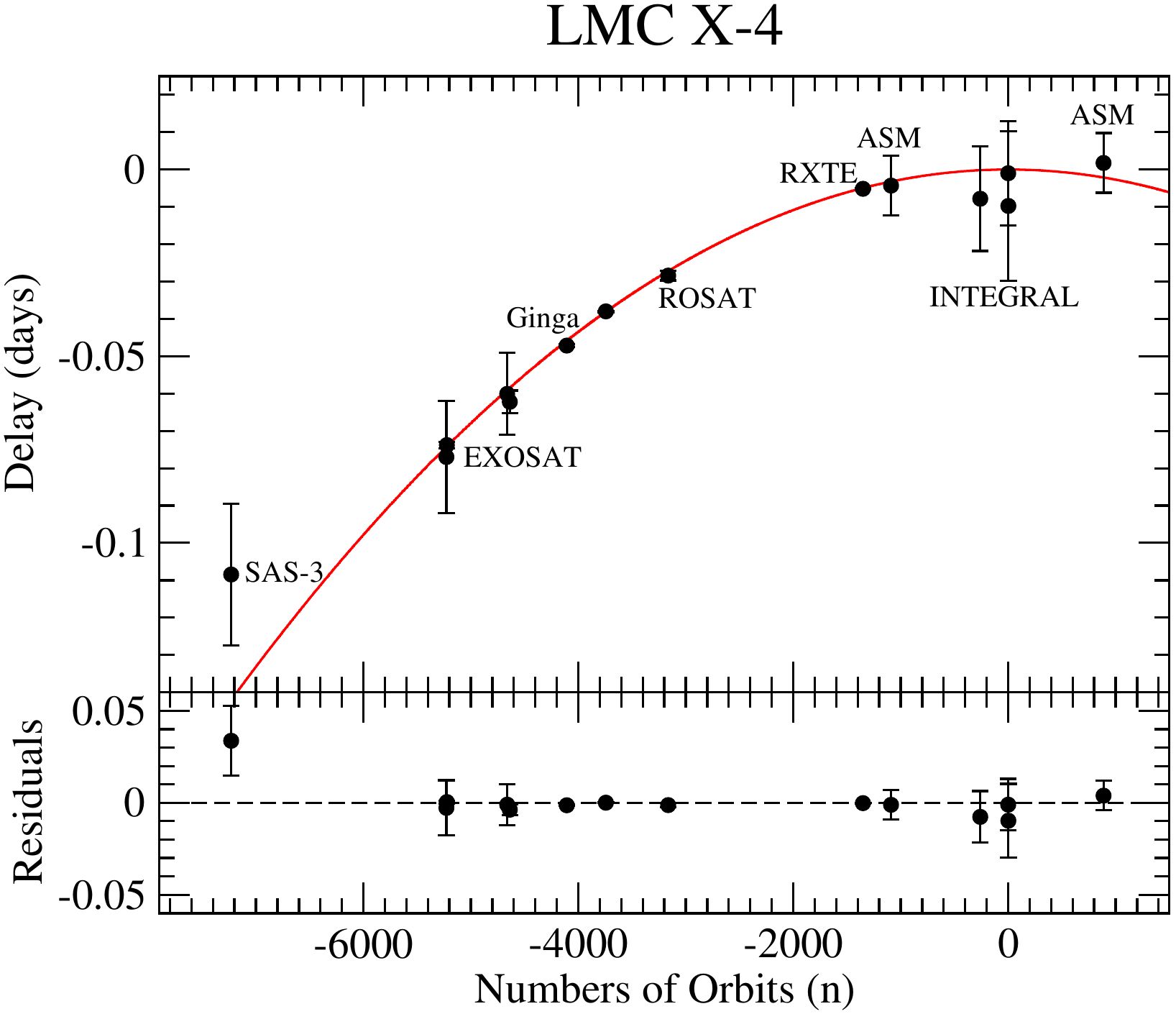}} & 
      \resizebox{55mm}{!}{\includegraphics{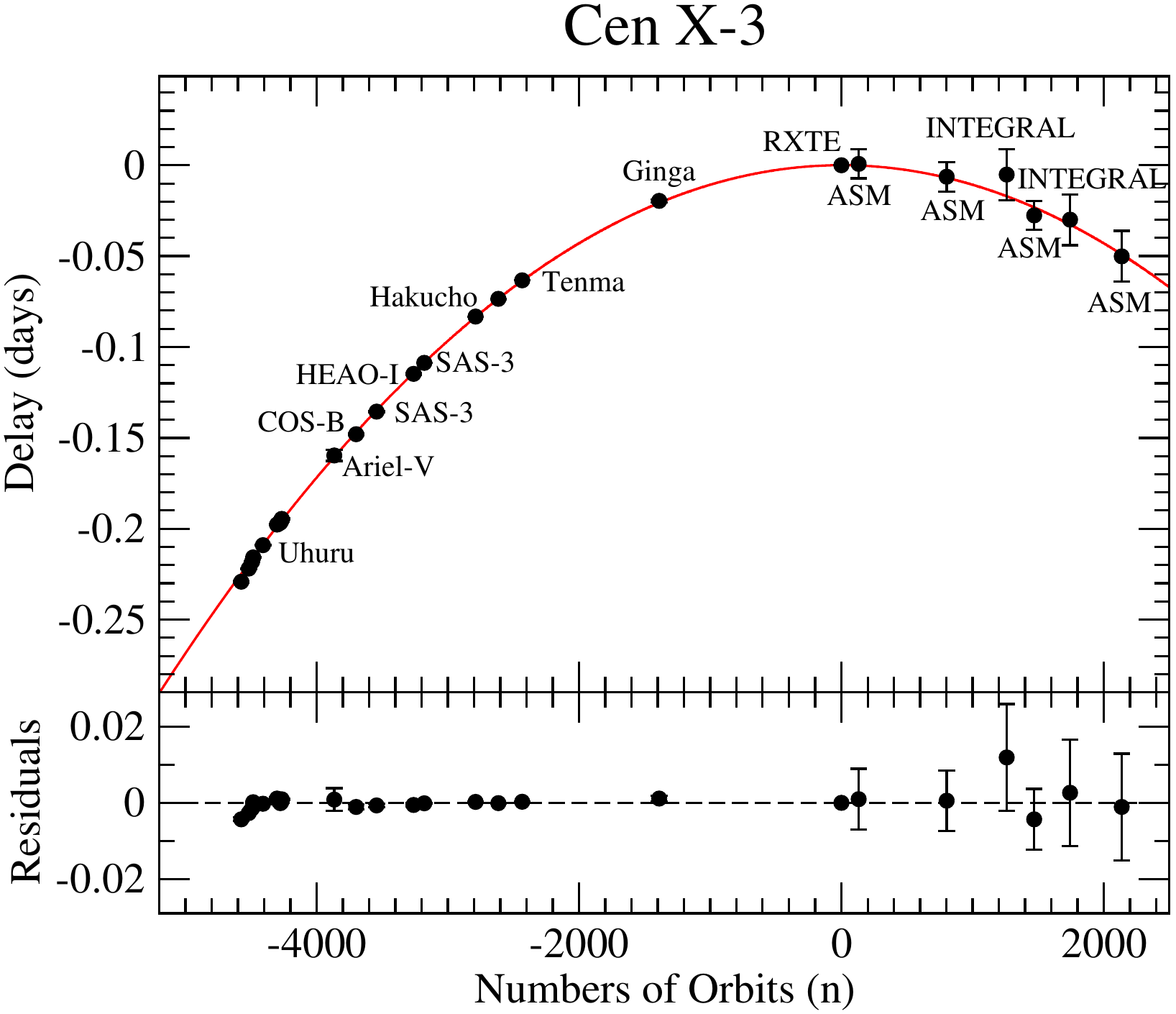}} &
      \resizebox{55mm}{!}{\includegraphics{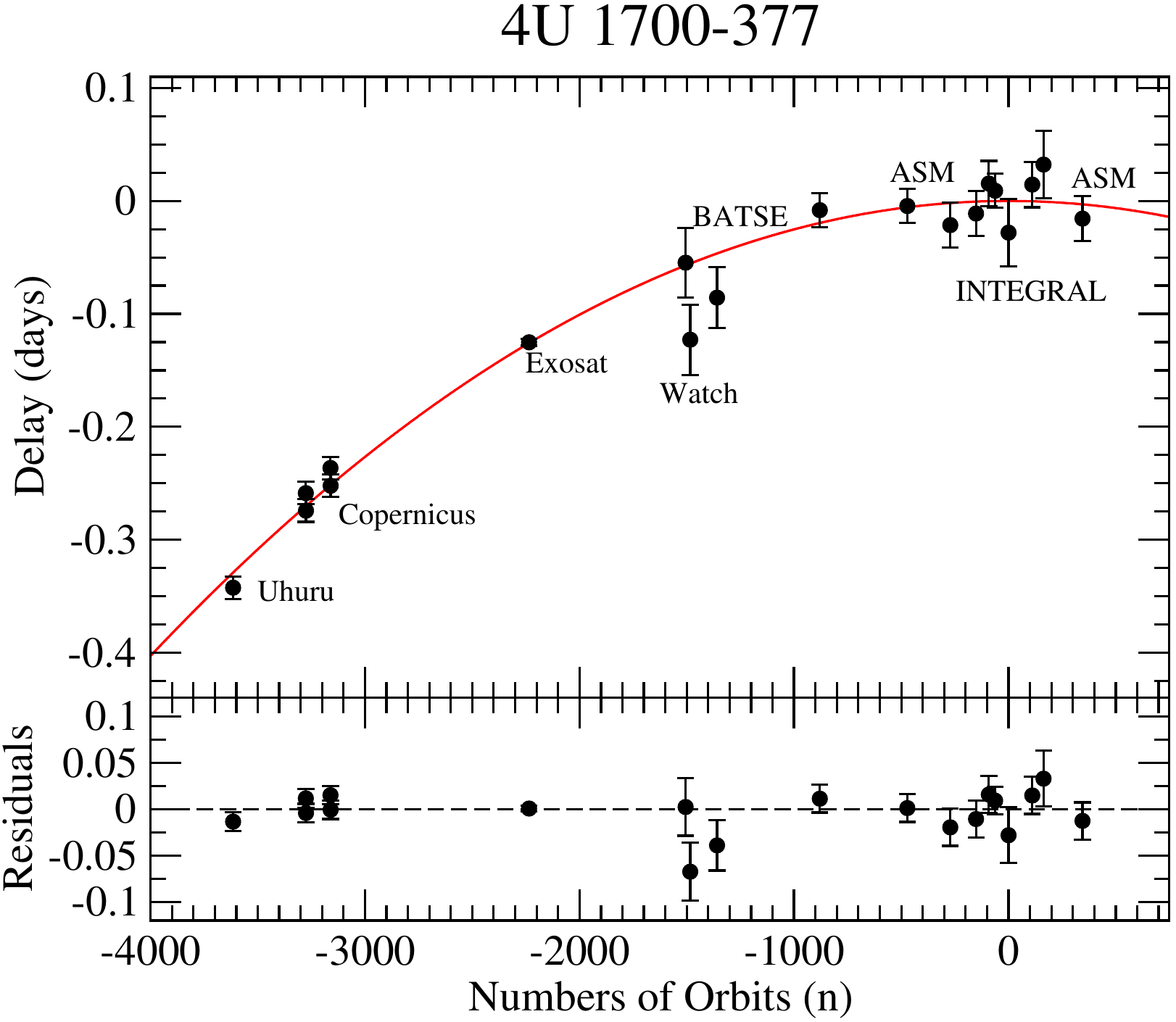}} \\
      \resizebox{55mm}{!}{\includegraphics{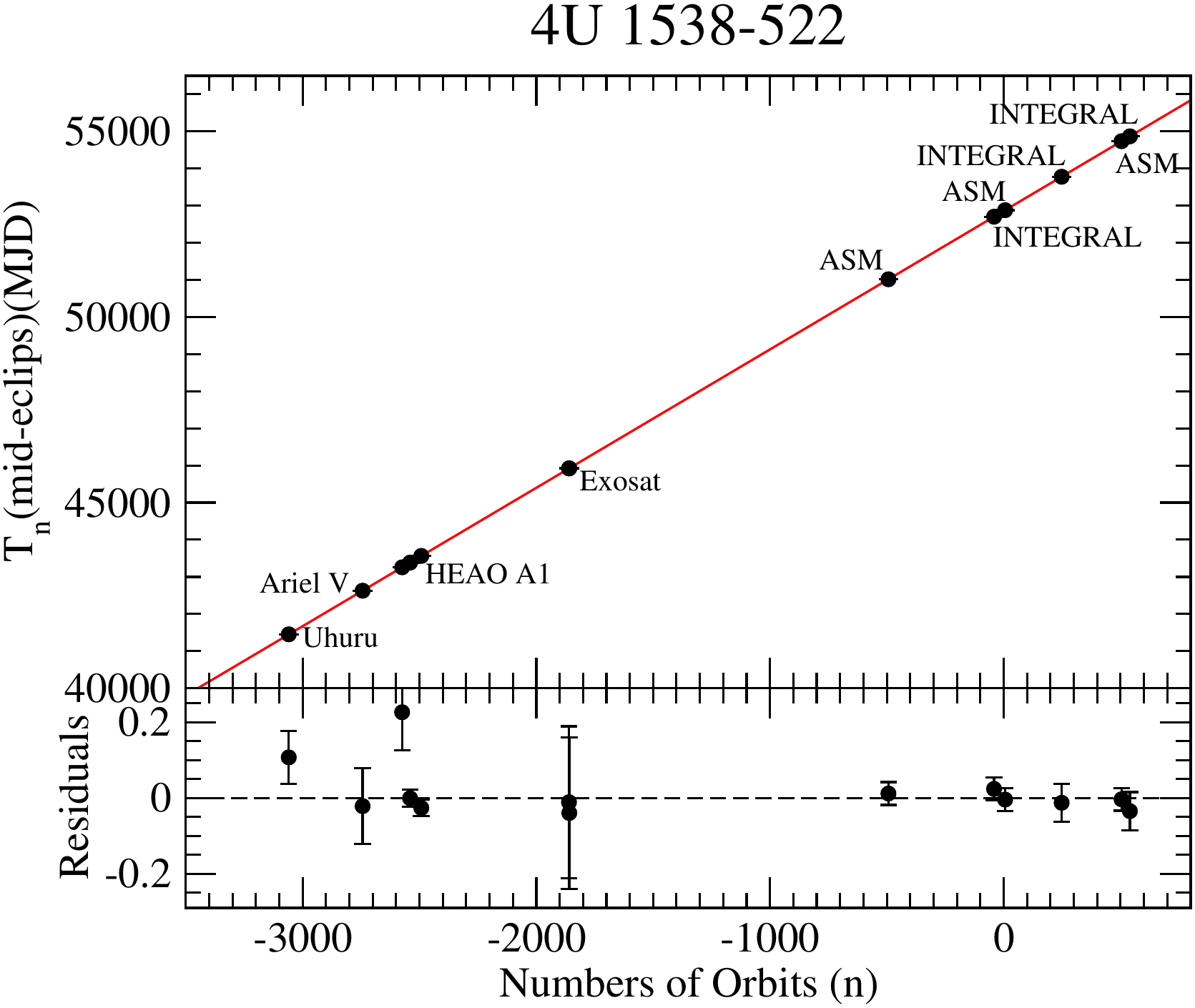}} &
      \resizebox{55mm}{!}{\includegraphics{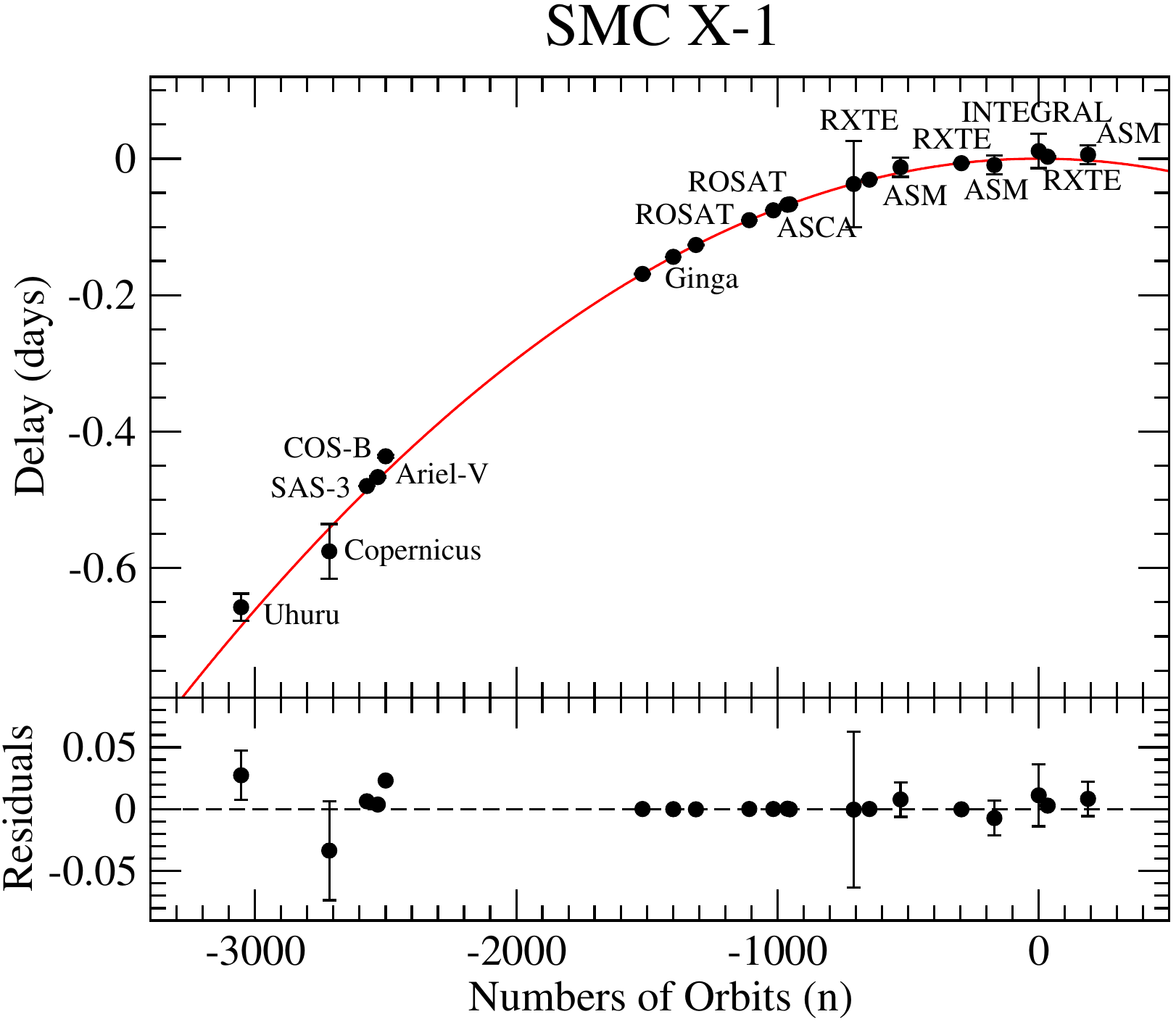}} &
      \resizebox{55mm}{!}{\includegraphics{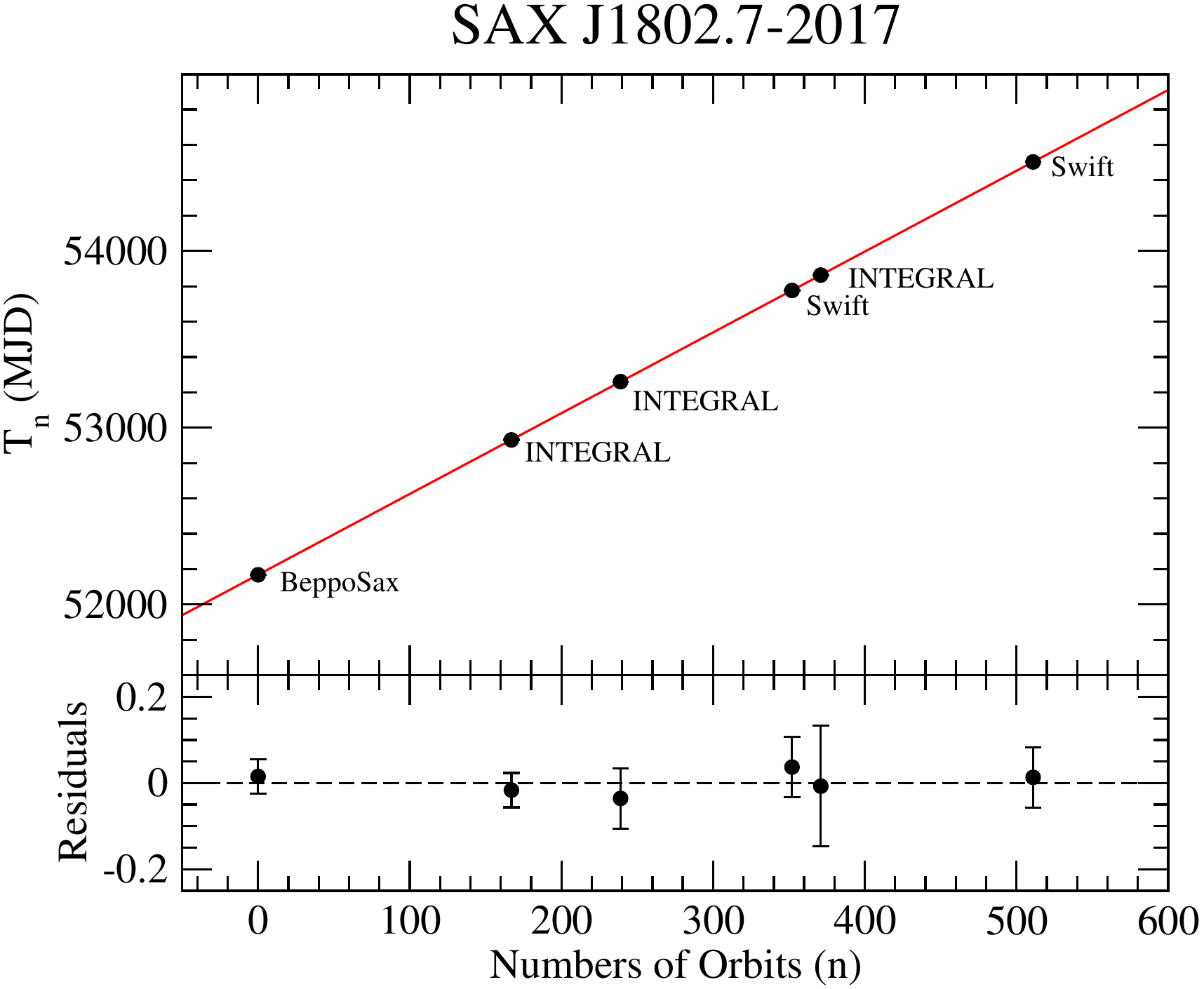}} \\
       \resizebox{55mm}{!}{\includegraphics{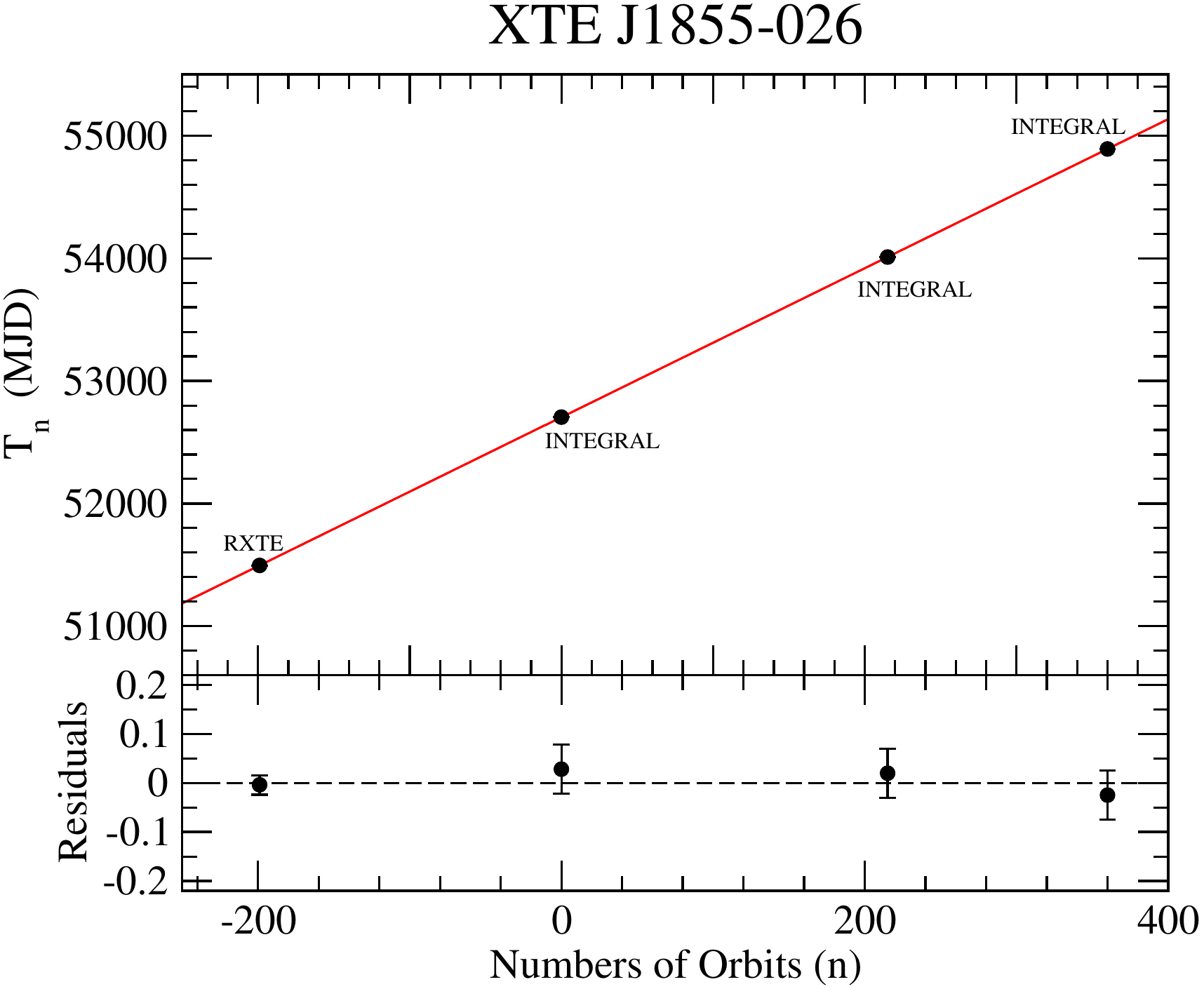}} &
      \resizebox{55mm}{!}{\includegraphics{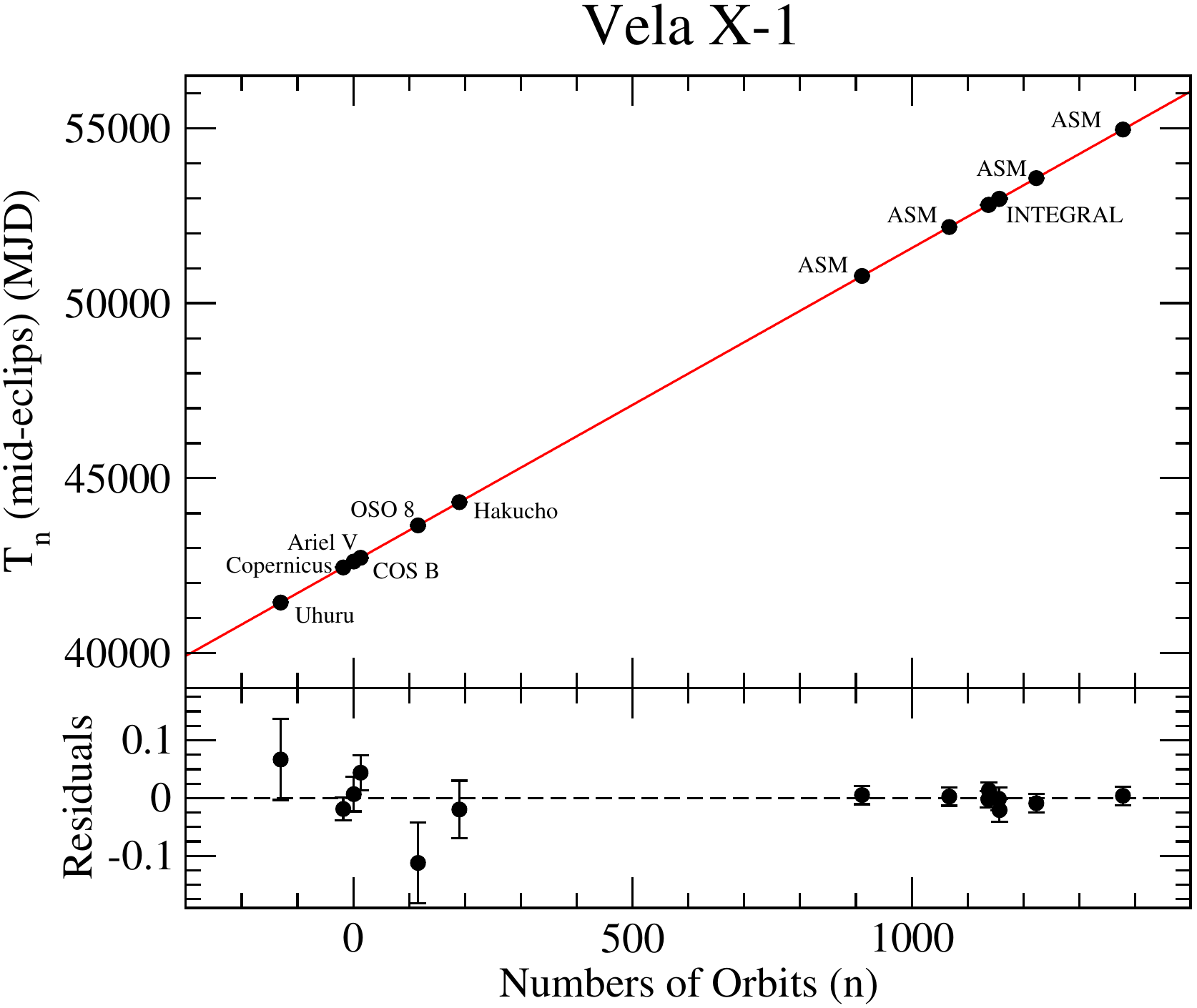}} &
      \resizebox{55mm}{!}{\includegraphics{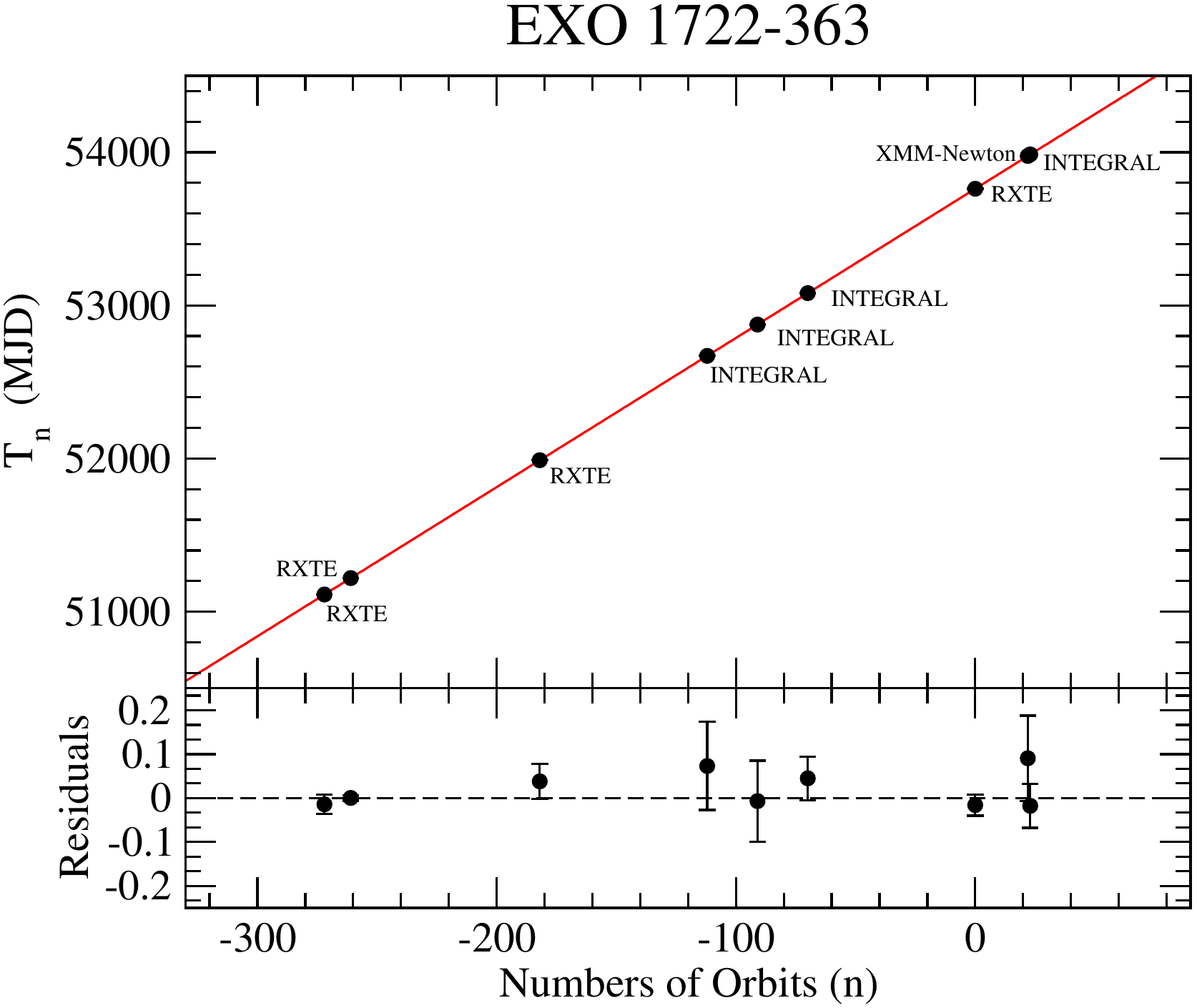}} \\
      \resizebox{55mm}{!}{\includegraphics{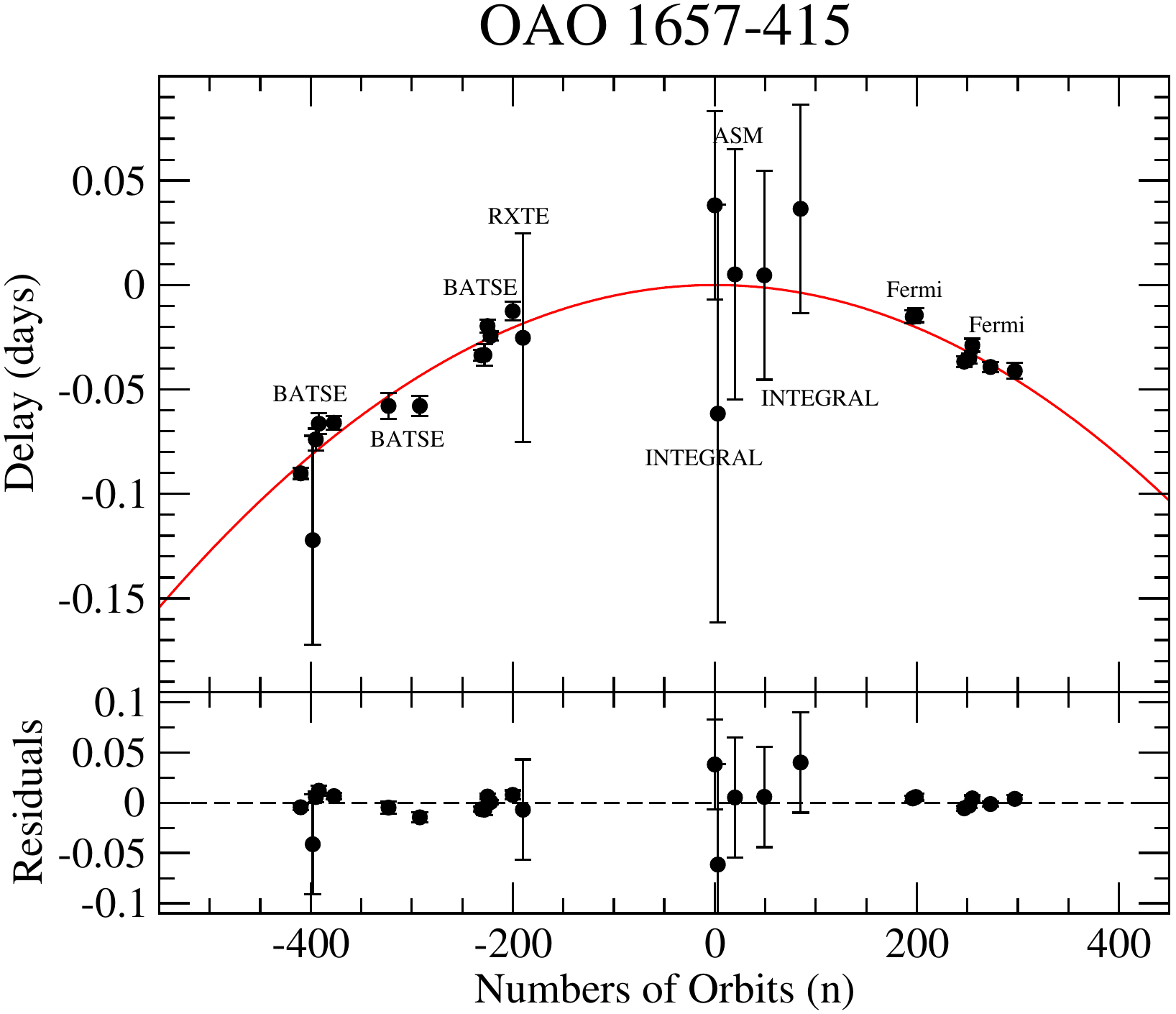}} &
   \end{tabular}
    \caption{All measured mid-eclipses, $T_{\rm ecl}$, mean longitudes, $T_{\rm \pi/2}$, and orbital epochs are shown with the best fit 
    models as a function of the orbit numbers. In all cases, the best fit model is a quadratic or linear fit to the epochs. 
    The lower panel in each figure shows the residual from the best fit.}
    \label{fig:orbits}
  \end{center}
\end{figure*}

For each of the selected sources we first determined the mid-eclipse (superior conjunction) times, $T_{\rm ecl}$, of all the eclipses 
found in the ISGRI and ASM data by using the e-fold method. We then improved these values by fitting together the newly determined 
epochs together with those derived from earlier observations (T$_{\rm n}$; see Appendix~\ref{table:ephemerides} and references 
therein) using the quadratic orbital change function:  
\begin{equation}
T_{n} = T_{\rm 0} + nP_{\rm orb} + \frac{1}{2}n^{2}P_{\rm orb}\dot P_{\rm orb}. 
\label{eq:1}
\end{equation}
Here $P_{\rm orb}$ is the orbital period in days, $\dot P_{\rm orb}$ is the period derivative at the epoch  $T_{\rm 0}$ and $n$ is the 
integer number of elapsed binary orbits. All the new measured orbital ephemeris are reported in Table~\ref{table:results}.  
We show in Fig.~\ref{fig:orbits} the best fit models to the data obtained by assuming either a simple linear orbital evolution or including 
a quadratic orbital decay term (the residuals from these fits are also shown).
For all those systems characterized by a non-negligible eccentricity ($\gtrsim0.1$), we fitted separately the mid-eclipse epochs, $T_{\rm ecl}$, 
and the mean longitude, $T_{\pi/2}$, epochs. For OAO~1657-415 we fit together the epochs of $T_{\rm ecl}$ and $T_{\pi/2}$ 
as the periastron angle of the source is known to be around 90$^{\circ}$ and therefore $T_{\rm ecl} \approx T_{\pi/2}$ (see Sec. \ref{sec:apsidal}). 
All the least square fits in this work were  performed by using the IDL tool \emph{MPFIT} \citep{Markwardt2009}. 
We obtained in all cases $\chi^{2}_{\rm red}\sim1.0-1.3$. 

In the case of the five sources LMC~X-4, Cen~X-3, 4U~1700-377, SMC~X-1, and OAO1657-415 we measured a significant orbital period derivative 
(see Table~\ref{table:results}). For 4U~1538-522, the $\dot{P_{\rm orb}}$ value has been obtained excluding the mid-eclipse times derived 
from the {\em Uhuru} and {\em Ariel} lightcurves \citep{Cominsky1991,Davison1977}; in the case of OAO~1657-415 we excluded the orbital epoch time 
reported by \citet{Barnstedt2008}. All these values were affected by relatively large uncertainties compared to the others and did not provide 
significant improvements to the fits. For the five sources characterized by a value of $\dot{P_{\rm orb}}$ consistent with zero 
(see Table~\ref{table:results}), the reported orbital epochs and periods were determined by using the linear orbital change function 
(see Eq.~\ref{eq:1}).

\subsection{Apsidal advance}
\label{sec:apsidal}

In eccentric orbits the time of mid-eclipse, $T_{\rm ecl}$, determined from the eclipses in the X-ray lightcurves, and the time of mean 
longitude, $T_{\pi/2}$, determined through the pulse arrival times technique, do not coincide. In this case we distinguished the orbital 
period, $P_{\rm orb, \pi/2}$, defined as the time elapsed between two successive passages at the same mean-longitude $l=\frac{\pi}{2}$, and 
the eclipse period, $P_{\rm orb, ecl}$, defined as the difference between two successive mid-eclipse epochs. The mean orbital longitude 
$l$ can be expressed as $l=M+\omega$, where, $M$ is the mean anomaly and $\omega$ is the argument of periapsis. With the orbital 
relations, expressed in the first order of the eccentricity, the time delay between $T_{\rm ecl}$ and $T_{\pi/2}$ can be written 
as \citep[see e.g.,][and references therein]{vanderKlis1984}:
\be
T_{\pi/2} - T_{\rm ecl}=\frac{eP_{\rm orb, \pi/2}}{\pi}\cos\omega.
\label{eq:tpi2tecl}
\ee
For an eccentricity of $e\sim0.1$ and a periastron angle of $\omega\sim150{\degr}$ as in Vela X-1, this can
result in a lag of $0.3$ day. This periodic lag is zero for eccentric orbits if $\omega=\pm90{\degr}$ and maximal 
when $\omega=0$ or $180{\degr}$. If the periastron angle $\omega$ is constant, then $P_{\rm orb, ecl}=P_{\rm orb, \pi/2}$. 
However, if the periastron is moving at a rate $\dot{\omega}$, the difference between $P_{\rm orb, ecl}$ and $P_{\rm orb, \pi/2}$
at the first order of its eccentricity, $e$, can be obtained by differentiating Eq.~\ref{eq:tpi2tecl} \citep[see][]{Deeter1987b}:
\be
P_{\rm orb, \pi/2}=P_{\rm orb, ecl}-\frac{eP_{orb, \pi/2}^{2}}{\pi}\dot{\omega}\sin\omega.
\label{eq:omegadot}
\ee
Equations~\ref{eq:tpi2tecl} and \ref{eq:omegadot} allow estimating $\omega$ and $\dot{\omega}$
if $\sin\omega\neq0$. For circular orbits we have $T_{\rm ecl} = T_{\pi/2}$ and thus $P_{\rm orb, ecl}=P_{\rm orb, \pi/2}$.
From Eq. (\ref{eq:omegadot}) and the value of $\Delta P_{\rm obs}= P_{\rm orb,ecl}-P_{\rm orb,\pi/2}$ determined from the observations, 
we calculated also the apsidal motion $\dot\omega$. The same epoch time is used to calculate both  
$(T_{\rm ecl},P_{\rm orb, ecl})$ and $(T_{\pi/2},P_{orb, \pi/2})$, as this time corresponds to the epoch of $\omega$ (see Table \ref{table1:sources}). 

Among the four sources with $e\gtrsim0.04$ considered in the present work (see Table~\ref{table1:sources}) a detailed comparison of 
the measured epoch time lag $\Delta T_{\rm obs}= T_{\rm \pi/2}-T_{\rm ecl}$, obtained from our best fit 
ephemerides with the predicted lag, with $\Delta T_{\rm calc}=eP_{\rm orb, \pi/2}\cos\omega/\pi$ could not be carried out for  
OAO~1657-415 (its periastron angle is $\omega \approx 90{\degr}$ and thus $T_{\rm ecl} \simeq T_{\pi/2}$), XTE~J1855-026 
(only one $T_{\pi/2}$ and $P_{\rm orb,\pi/2}$ are available to date; see Appendix \ref{table:ephemerides}) and EXO~1722-363 
\citep[no reliable eccentricity measurement is available in the literature;][]{Thompson2007}.
For 4U~1538-522 we used the eccentricity, $e$, and the periastron angle, $\omega$, from \citet{Clark2000} and \citet{Mukherjee2006};  
for Vela X-1 we used the ephemeris and orbital parameters from \citet[][see also Table~\ref{table1:sources}]{Rappaport1976}. For both sources 
the best fits are obtained using the linear orbital change function (Eq.~\ref{eq:1}). All results are reported in Table \ref{table:apsidal}. 
In the case of 4U~1538-522 two different fits are reported. In the first, we fit all the mean longitude epochs ($T_{\pi/2}$) 
until 1997 with the last one reported by \citet{Clark2000}; in the second, we included all the longitude epochs until 2003, 
with the last one reported by \citet{Mukherjee2006}. The two cases give compatible results (to within the uncertainties) only if the epochs 
reported from \citet{Clark2000} are excluded from the second fit. Note that the same mid-eclipse time, $T_{\rm ecl}$, has been used 
for both fits (see Table \ref{table:apsidal}). 
\begin{table*}
\caption{Updated epochs, orbital periods, and period decay of the ten sources obtained by using all the available mid-eclipse times. 
We indicated in brackets the uncertainties at $1\sigma$ c.l. on the last digits of each reported value.}
\begin{center}
\begin{tabular}{lllll}
\hline 
Source & $T_{\rm 0,ecl}$ (MJD) &  $P_{\rm orb,ecl}$ (days) & $\dot{P}_{\rm orb}/P_{\rm orb}$ (10$^{-6}$yr$^{-1}$) & $\dot \omega$ (deg/yr)\\
\hline
\hline 
LMC X-4 & 53013.5910(8) &1.4083790(7) &$-1.00(5)$ & --\\
Cen X-3 &  50506.788423(7) & 2.08704106(3)& $-1.800(1)$ & --\\
4U 1700-377 & 53785.850(7)  &3.411581(7) & $-1.6(1)$ &-- \\
4U 1538-522$^{a,b}$ & 52855.061(13) &3.7284140(76) & $-0.7(8)$ &1.3(6)\\
SMC X-1 & 52846.688810(24) & 3.891923160(66)& $-3.541(2)$ & --\\
SAX J1802.7-2017$^{b}$ & 52168.245(34) & 4.5697(1) & $17(29)$ & --\\
XTE J1855-026$^{b}$ &  52704.009(17) & 6.07415(8) &  $-12(13)$ & --\\
Vela X-1$^{b}$ & 42611.349(13) & 8.964427(12) & $-0.1(3)$ &0.41(27)\\
EXO 1722-363$^{b}$ & 53761.695(19) & 9.74079(8) &  $-21(14)$ &-- \\
OAO 1657-415 & 52674.1199(17) & 10.447355(92) &  $-3.4(1)$ &--\\
\hline
\end{tabular}
\end{center}
\tablefoot{$^{a}$Epoch time and orbital period is derived from the linear fit; the orbital decay has been  
evaluated excluding the {\em Uhuru} and {\em Ariel} data. $^{b}$ The orbital period derivative, $\dot P$ are consistent with zero 
and thus the epoch times and orbital periods are estimated using Eq.~\ref{eq:1} (linear fit).}
\label{table:results}
\end{table*}
\begin{table*}
\tiny
\caption{In this table, the predicted $\Delta T_{\rm calc}$ (Eq.~\ref{eq:tpi2tecl}), the measured epoch time lag, 
$\Delta T_{\rm obs}= T_{\rm \pi/2}-T_{\rm ecl}$, and the orbital period lag, $\Delta P_{\rm obs}= P_{\rm orb,ecl}-P_{\rm orb,\pi/2}$ are given 
for the two sources 4U 1538-522 and Vela X-1. The apsidal advance angle, $\dot\omega$, is also reported. 
We indicated in brackets the uncertainties at $1\sigma$ c.l. on the last digits of each reported value.}
\begin{center}
\begin{tabular}{lcccccccc}
\hline 
Source & $T_{\rm ecl}$  & $T_{\rm \pi/2}$ & $P_{\rm orb,ecl}$ & $P_{\rm orb,\pi/2}$ & $\Delta T_{\rm obs}$  & 
$\Delta T_{\rm calc}$& $\Delta P_{\rm obs}$ & $\dot \omega$ \\
& (MJD)& (MJD)& (day)&(day) & (day)& (day)& ($\times10^{-4}$ day)&(deg/yr)\\
\hline
\hline 
4U 1538-522$^{a}$ & 50450.234(11) & 50450.221(11) & 3.7284140(76)& 3.728337(22) &-0.013(15) & 0.0905(69) &0.72(23)&2.3(7)\\
4U 1538-522$^{b}$ &  52851.332(13) & 52852.3207(92)&3.7284140(76) & 3.728382(11)&0.011(16)&0.164(67)&0.32(13)&1.3(6)\\
Vela X-1 & 42611.349(13)& 42611.1693(43)&8.964427(12) & 8.9644061(64) & -0.180(14) & -0.2275(55) & 0.21(14)&0.41(27)\\
\hline
\end{tabular}
\end{center}
\tablefoot{$^{a}$The best fit values $T_{\pi/2}$ 
and $P_{\rm orb,\pi/2}$ were obtained including all the epochs available in the literature until the work of 
\citep[the last one is reported by][]{Clark2000}. $^{b}$Same as before but using all epochs available until the work 
published by \citet{Mukherjee2006}.}
\label{table:apsidal}
\end{table*}

\section{Folded lightcurves}
\label{sec:folded_lc}
 
\begin{figure*}[ht!]
  \begin{center}
   \begin{tabular}{ccc}
      \resizebox{55.8mm}{!}{\includegraphics{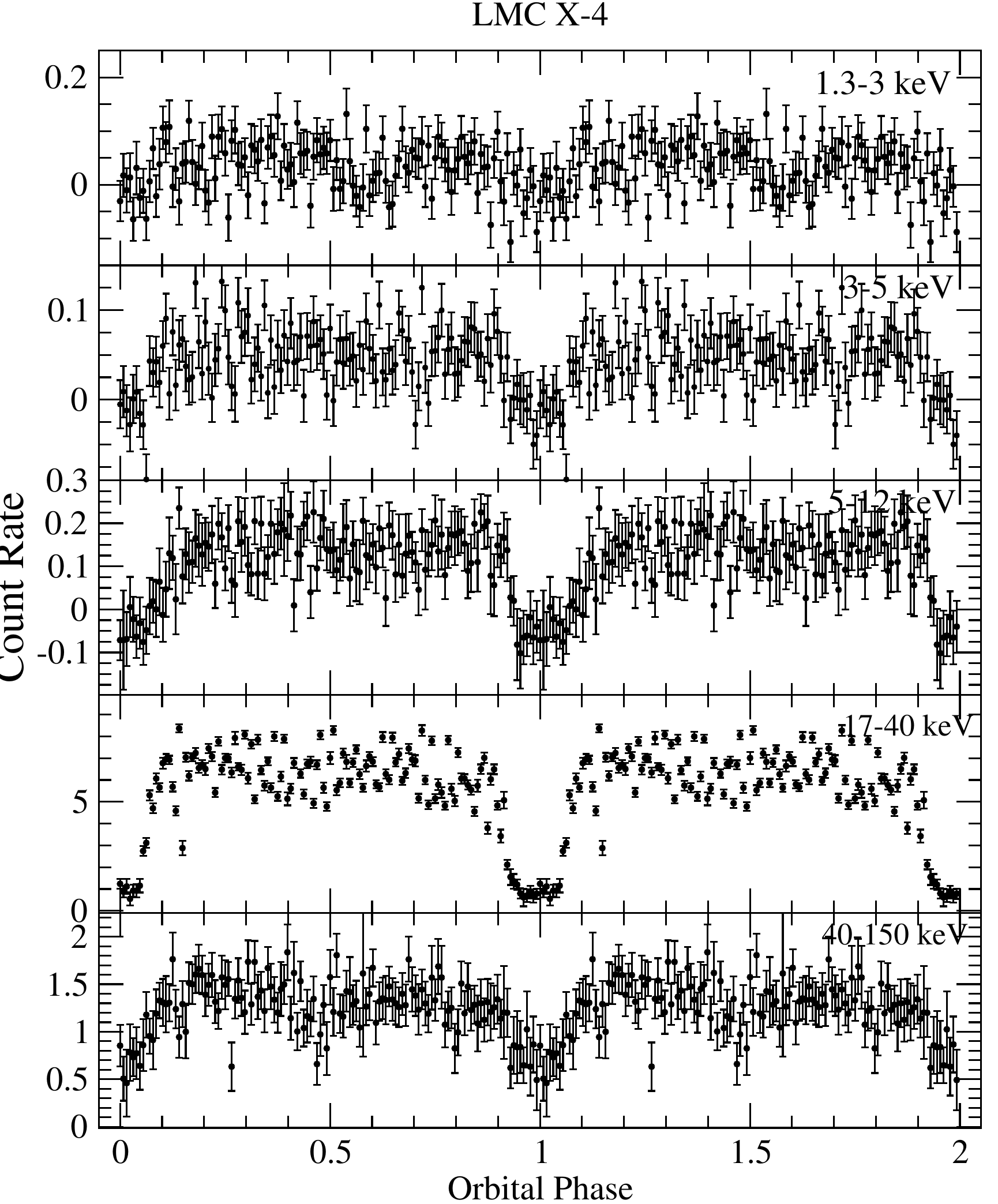}} & 
      \resizebox{53mm}{!}{\includegraphics{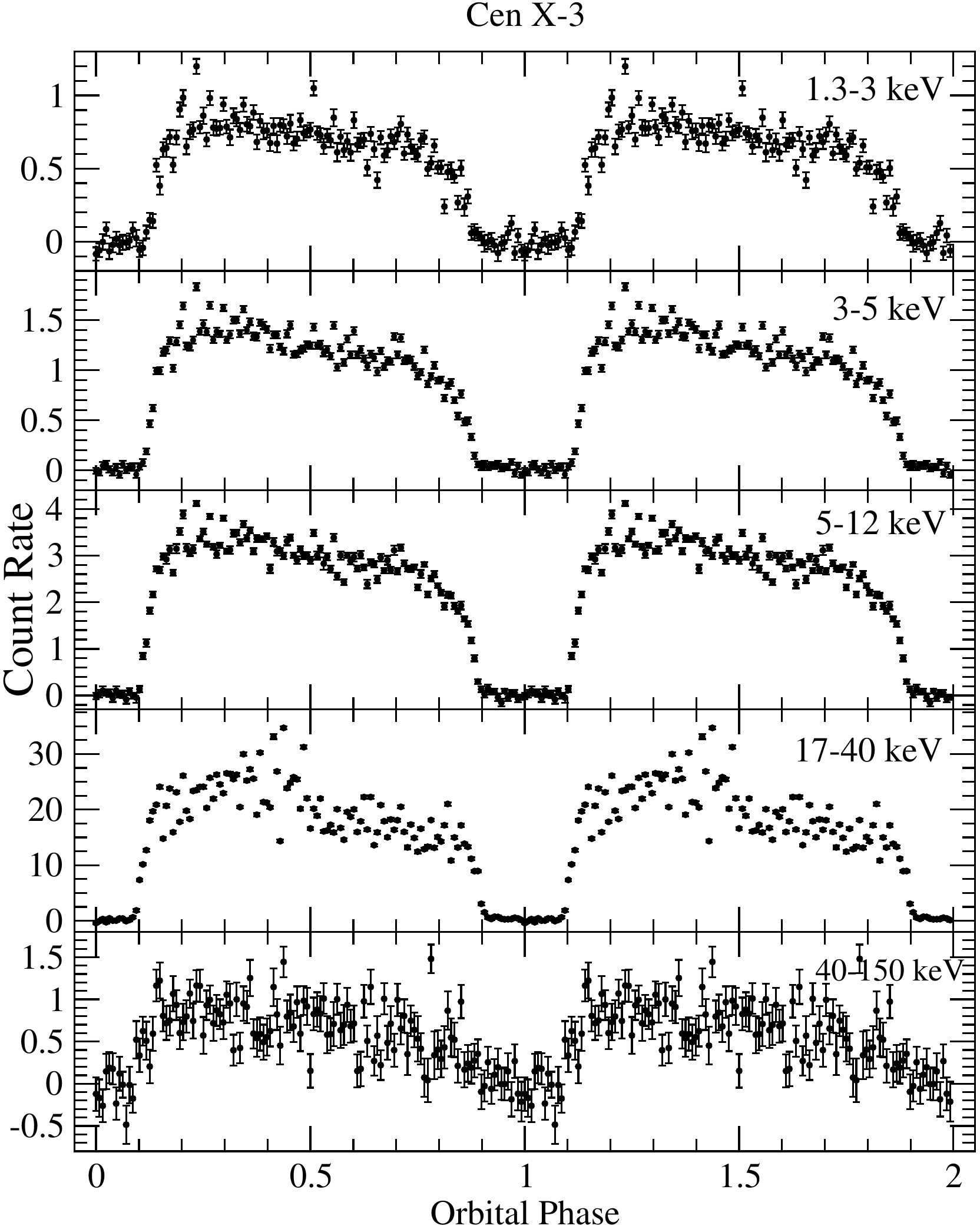}} &
     \resizebox{52.5mm}{!}{\includegraphics{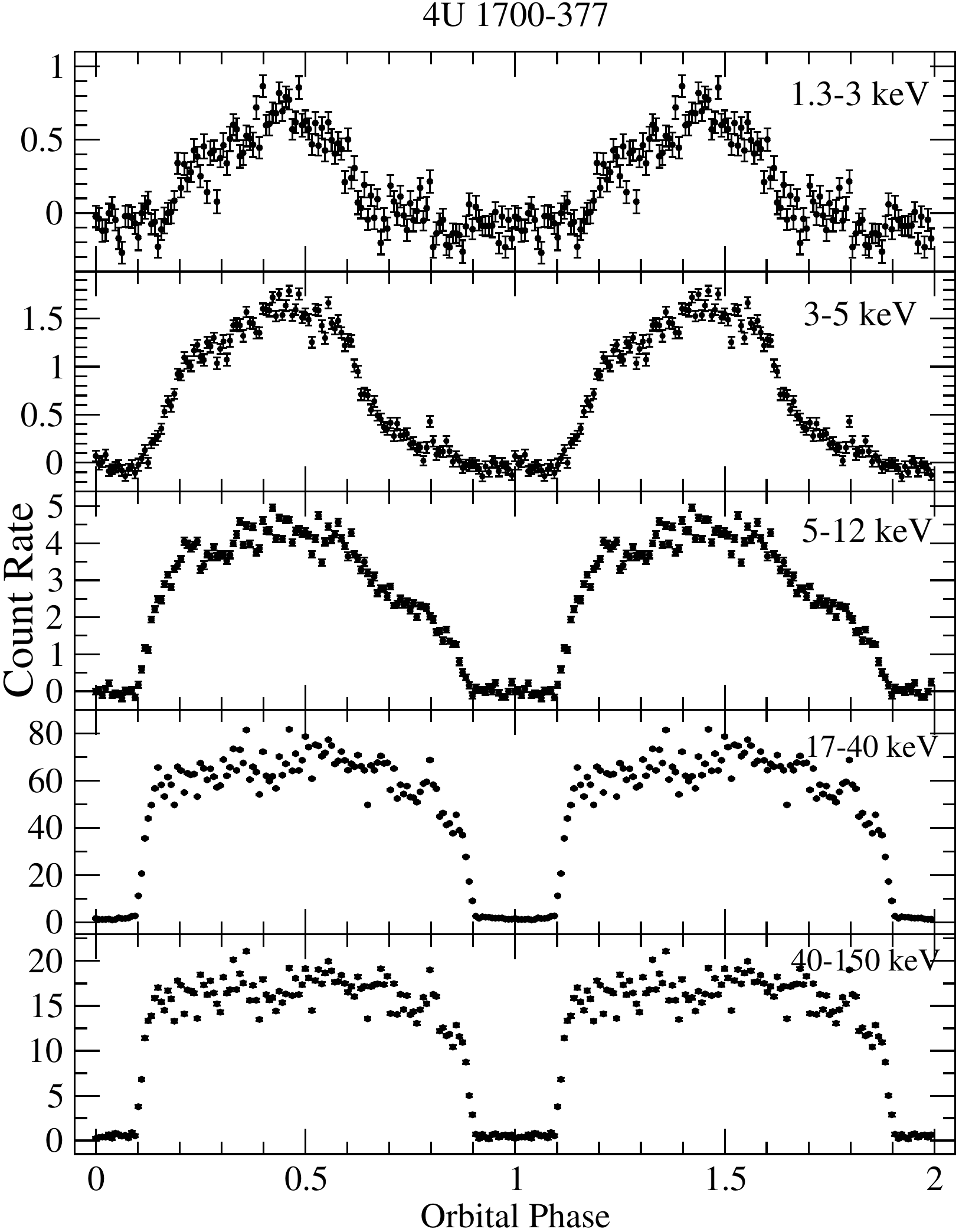}} \\
      \resizebox{56mm}{!}{\includegraphics{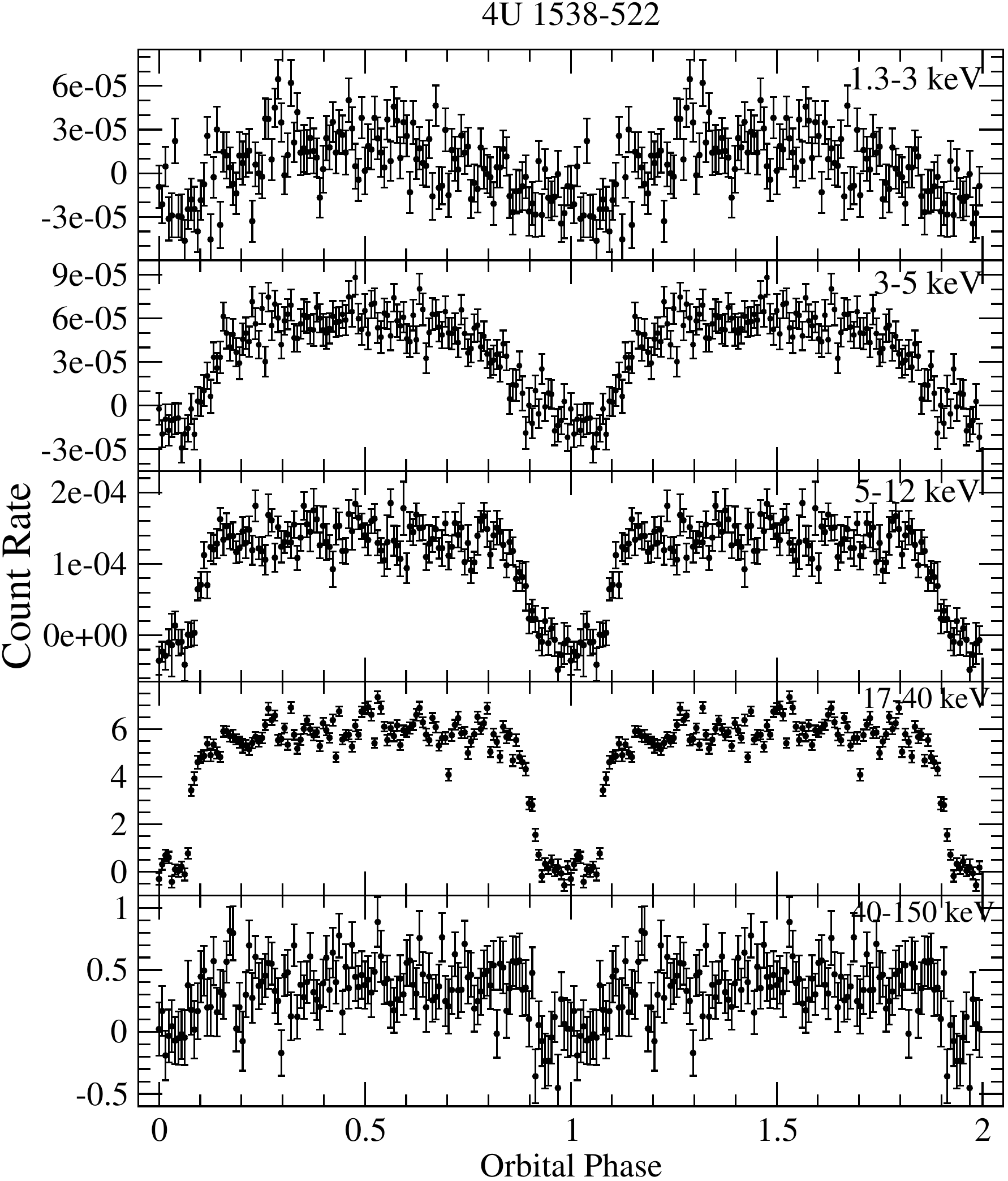}} &
      \resizebox{54mm}{!}{\includegraphics{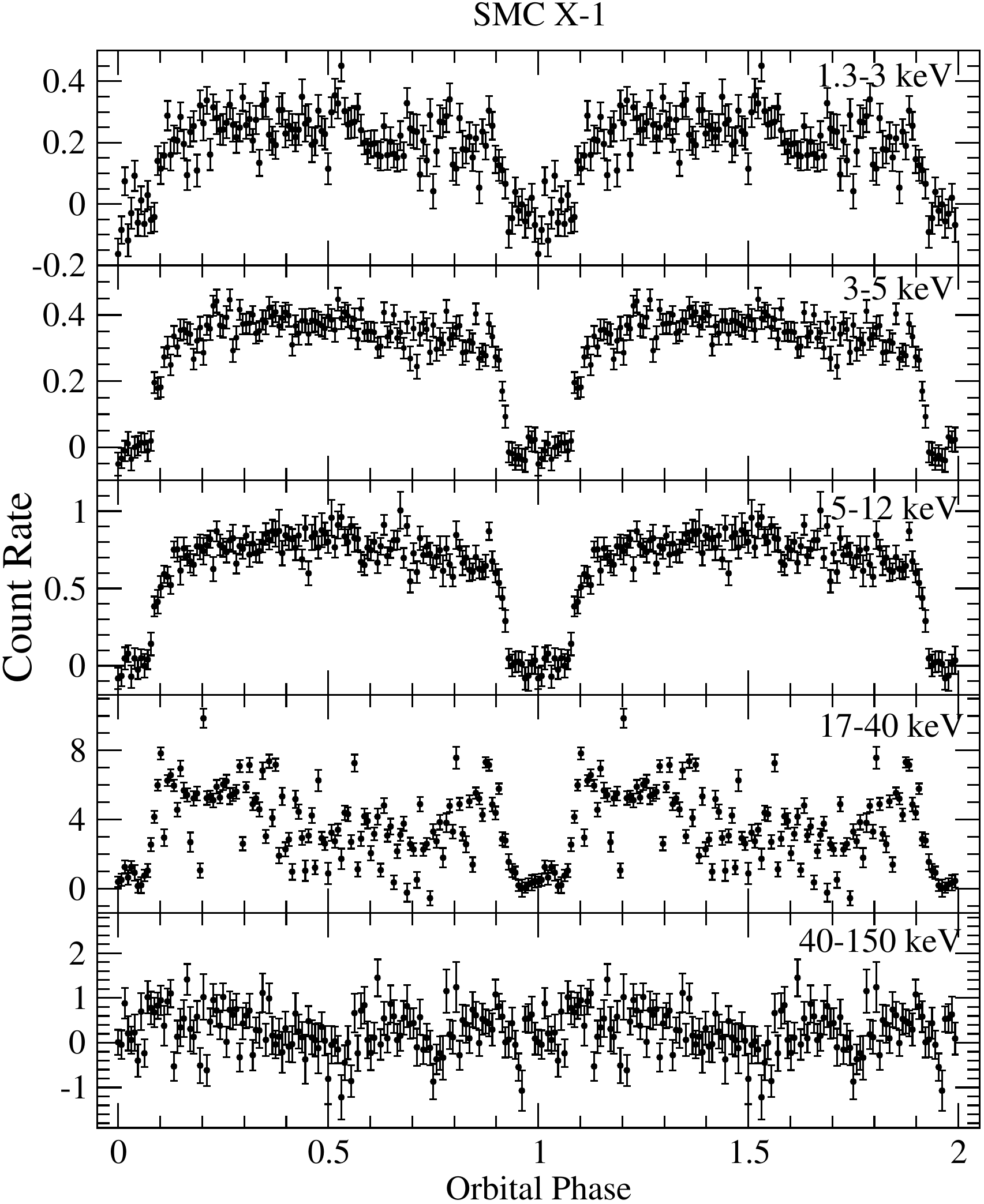}} &
      \resizebox{55mm}{!}{\includegraphics{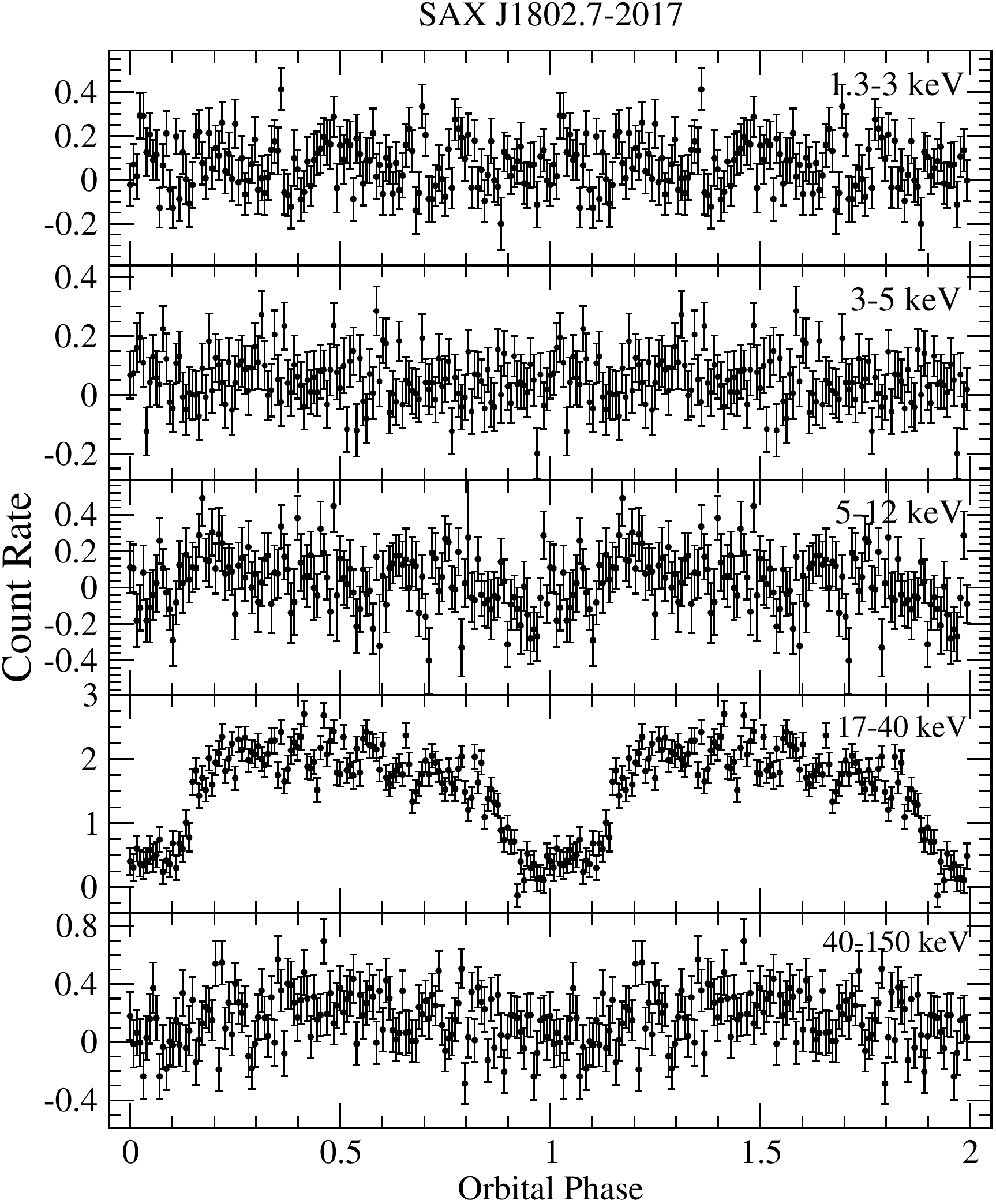}} \\
   \end{tabular}
    \caption{Folded lightcurves of the first 6 eclipsing HMXBs in Table \ref{table1:sources}. Data in the energy range 
    1.3-12 keV were obtained from the RXTE/ASM, while harder X-ray data are from INTEGRAL/ISGRI (17-150 keV). In 
    all cases, the lightcurves have been folded by using the updated ephemerides obtained in the present work.}
    \label{fig:lightcures}
  \end{center}
\end{figure*}

\begin{figure*}[ht!]
  \begin{center}
   \begin{tabular}{ccc}
           \resizebox{53mm}{!}{\includegraphics{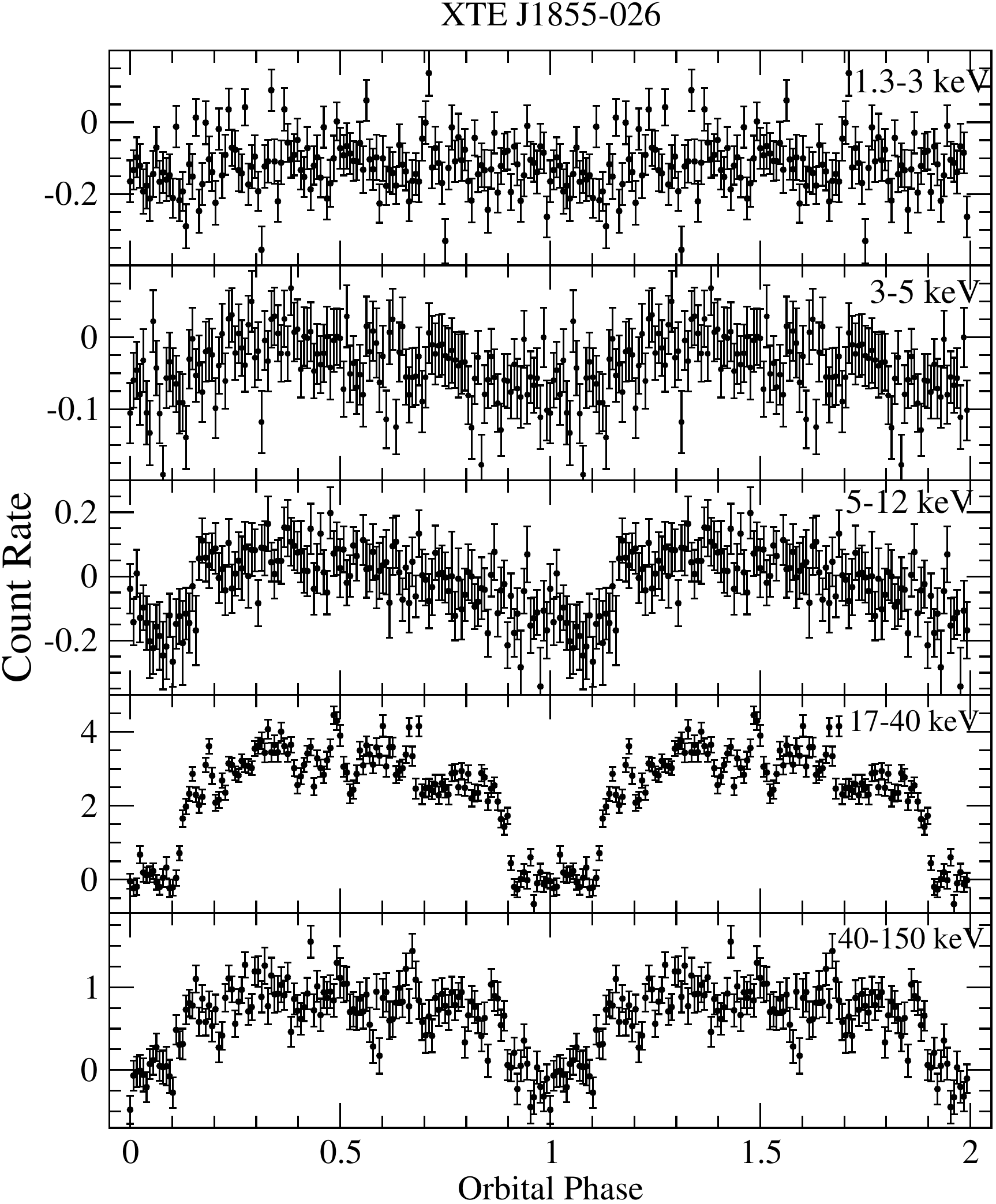}} &
     \resizebox{53mm}{!}{\includegraphics{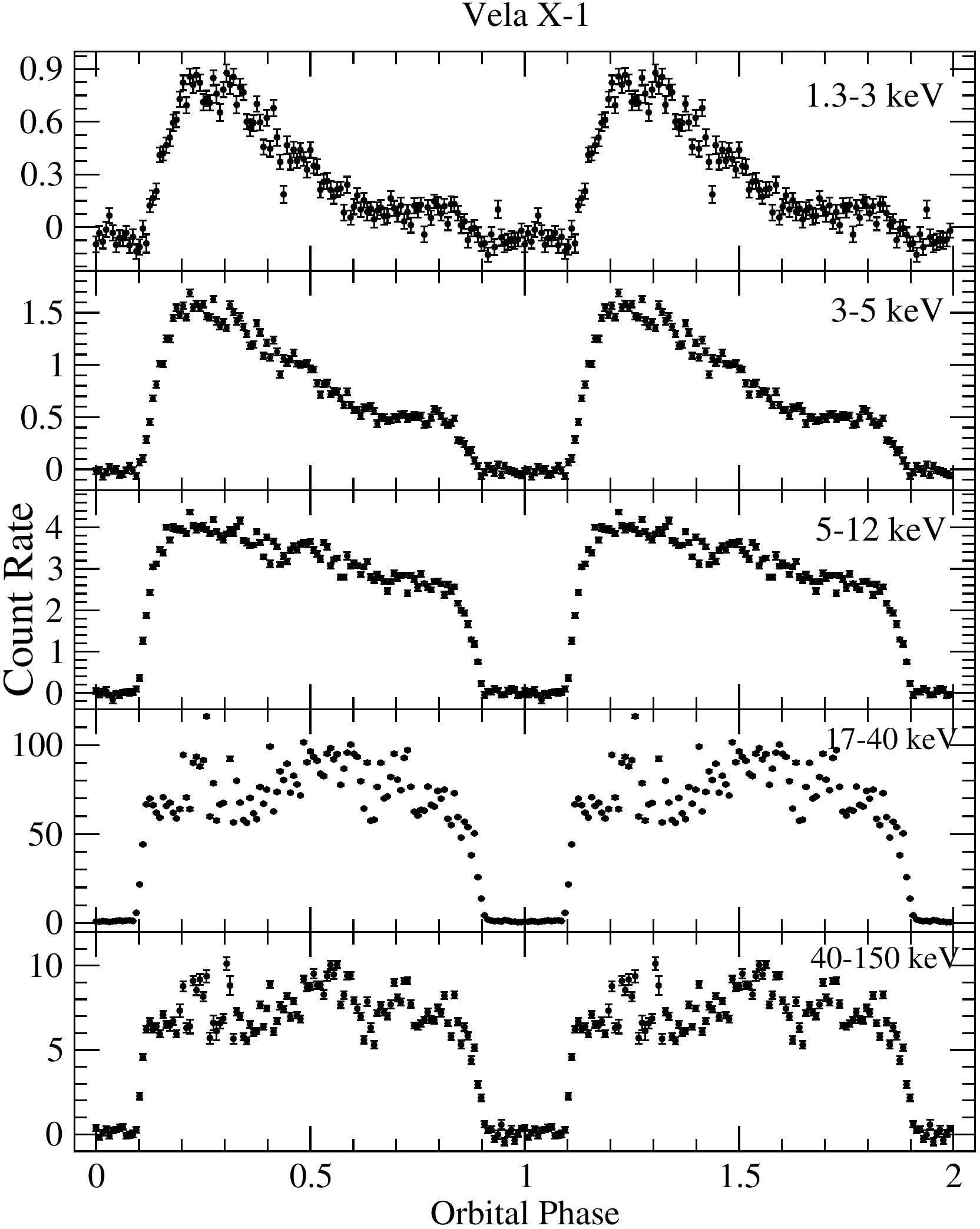}} \\
          \resizebox{55mm}{!}{\includegraphics{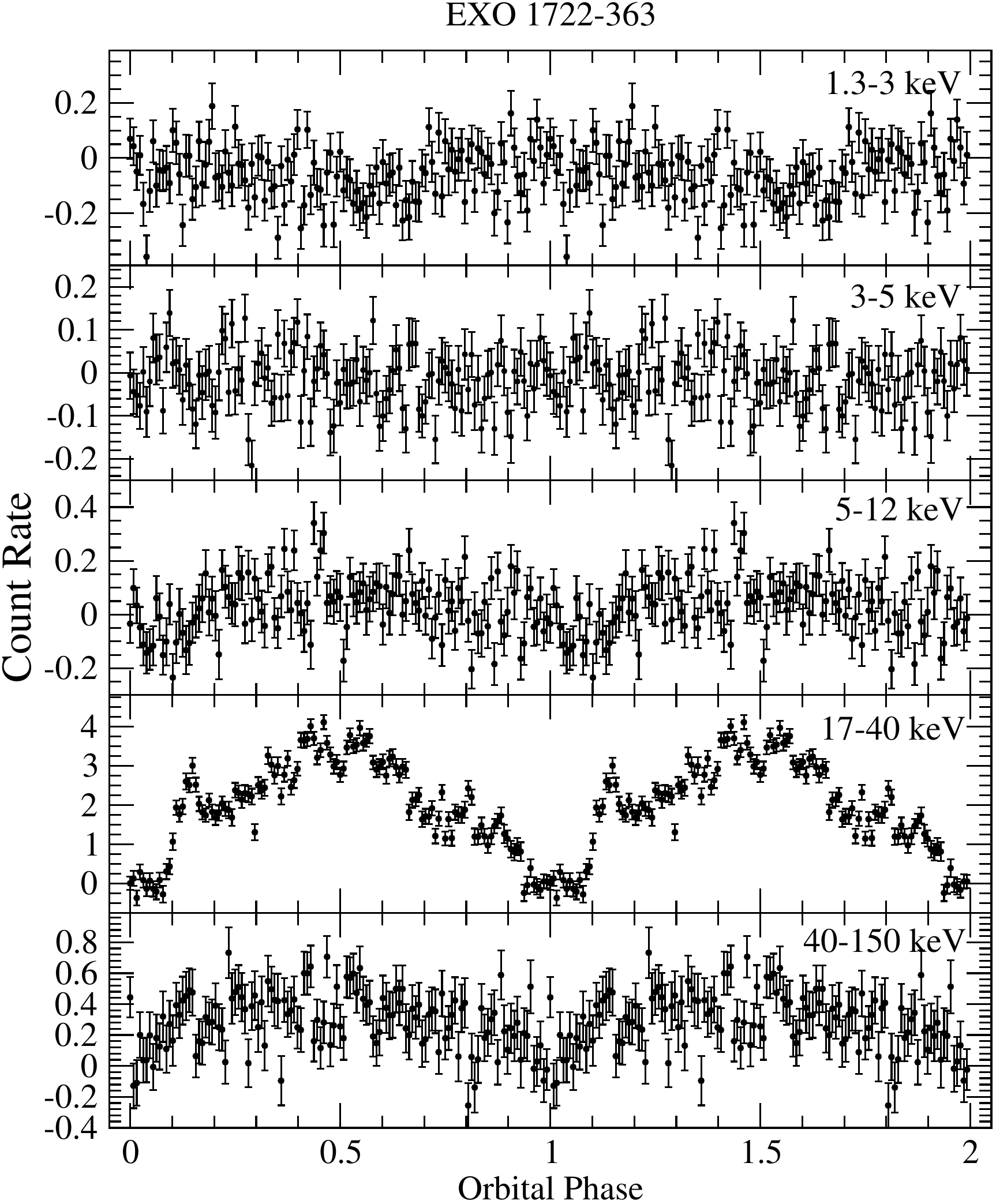}} & 
               \resizebox{54.5mm}{!}{\includegraphics{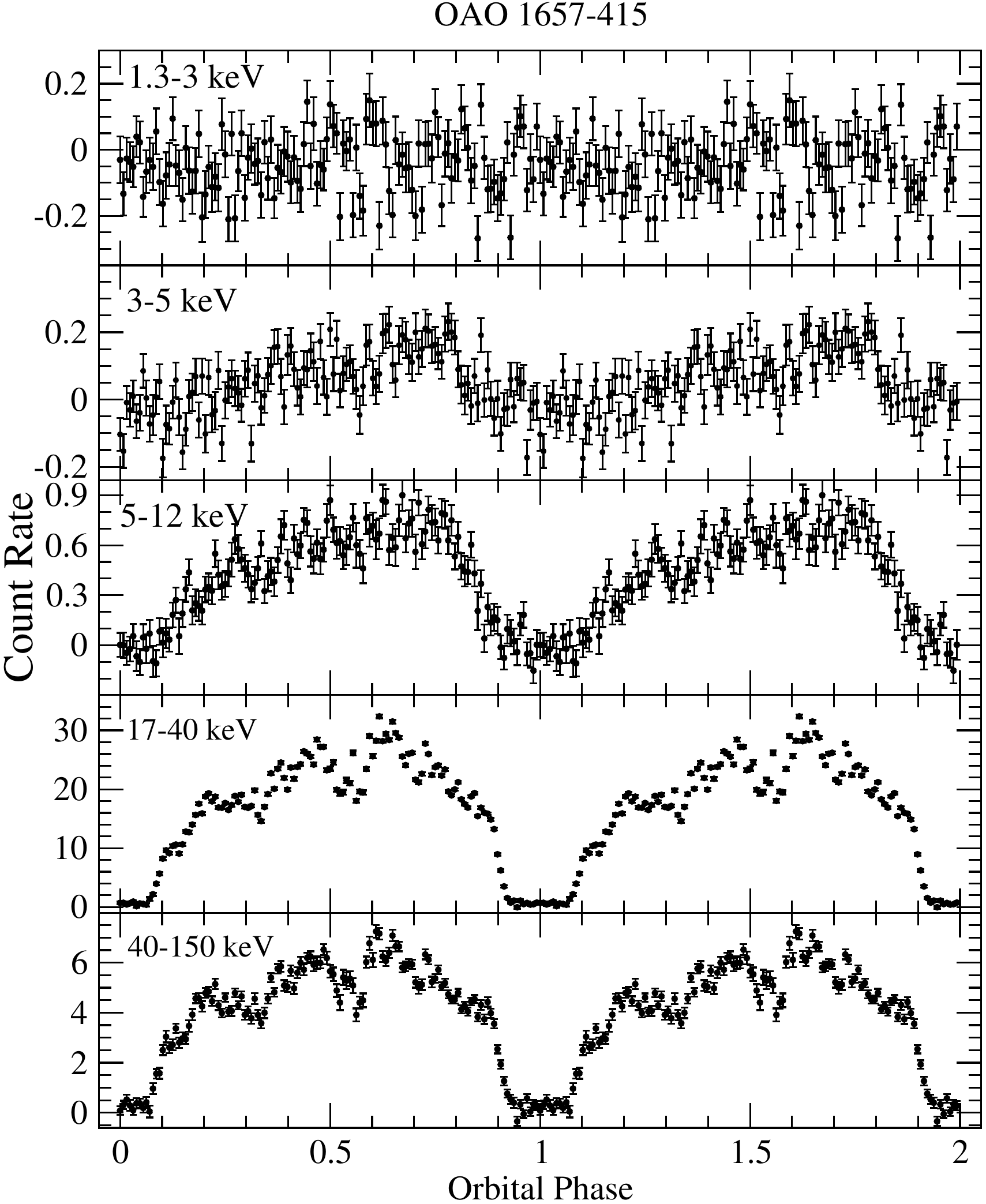}}&
   \end{tabular}
    \caption{Same as Fig. \ref{fig:lightcures} but for the remaining 4 sources in Table \ref{table1:sources}.}
    \label{fig:lightcures2}
  \end{center}
\end{figure*}

The updated ephemerides we obtained for the ten HMXBs allowed folding their lightcurves with an unprecedented accuracy. 
We folded for each source the dwells \rxte/ASM light curves  in the 1.3--3 keV, 3--5 keV, and 5--12 keV bands, 
and the \I/ISGRI light curves in the 17--40 keV and 40--150 keV energy bands by using 128 phase bins (see Fig.~\ref{fig:lightcures} and  \ref{fig:lightcures2}). 
For LMC~X-4, Cen~X-3, 4U~1700-377, SMC~X-1, and OAO~1657-415 the orbital period derivative was also taken into account 
during the folding. Given the values of $\dot{P}_{\rm orb}/P_{\rm orb}$ in Table \ref{table:results}, we obtained for these sources 
a maximum variation of the orbital period (between the first and last mid-eclipse time  of the \rxte/ASM dataset) 
of $\sim$ 0.088, 0.158, 0.139, 0.310, and 0.299~days, respectively. 
These delays have been calculated by using the quadratic term in Eq.~\ref{eq:1} rewritten as $(\Delta t)^{2}\dot{P}_{\rm orb}/{P}_{\rm orb}$ 
(we replaced $nP_{\rm orb}$ by the observational elapsed time, $\Delta t$). 
The orbital derivatives for these sources are not negligible, as the derived correction factors in time are larger than a phase bin ($1/128)$ 
and the inclusion of these corrections significantly improve the shape of the folded lightcurves.

The fluxes (cts/s) of the ten sources measured during the occultation of the compact object by the companion star are reported in 
Table~\ref{table:occultations_flux}. In all cases, the fluxes in the lower energy bands (1.3--3 keV, 3--5 keV, and 5--12 keV energy bands) 
are consistent with zero. This suggests that in the soft X-ray domain the source emission is strongly absorbed by the extended corona 
of the companion star, as well as from the companion star itself \citep[in the literature, residual fluxes have been 
reported at a level that is too low to be detectable by the ASM on-board RXTE; see, e.g.,][]{Ebisawa1996,Lutovinov2000,Vrtilek2001}. At higher energies (17--40 keV and 40--150 keV), the observed 
residual X-ray fluxes are most likely due to the presence of an extended X-ray scattering region (e.g., the accretion disc around the 
compact object, the X-ray irradiated surface of the companion star, or a cloud of material diffused around the system and produced by the intense wind of the supergiant star). These findings, together with the different shapes of the eclipse ingress and egress in 
Fig.~\ref{fig:lightcures} and  \ref{fig:lightcures2}, are discussed in Sect.~\ref{sec:eclipse}. 

\subsection{Semi-eclipse angles}
\label{sec:semi-eclipse}

We used the folded lightcurves in the energy band 17--40 keV to estimate the semi-eclipse angle, $\theta_{\rm e}$, 
of each eclipse observed from the ten selected sources in Table~\ref{table1:sources}. In this energy band, all eclipses 
look sharp and symmetric, thus permitting to achieve an unprecedented accuracy in the determination of the occultation time 
(see Fig. \ref{fig:lightcures} and \ref{fig:lightcures2}). We followed the fitting method described in \citet{Rubin1996}. 
The duration of each eclipse was calculated from the measured phase of the eclipse ingress and egress. 
\begin{table}[ht!]
\caption{Measured cts/s during the occultation phases of the ten selected sources in this work. 
Note that 0.2~cts/s (0.1~cts/s) in the 17--40~keV (40-150~keV) energy band in ISGRI 
correspond to roughly 1~mCrab. We indicated in brackets the uncertainties at 90\% c.l. on the last digits of each reported value.}
\begin{center}
\begin{tabular}{lll}      
\hline 
Source & 17 -- 40 keV & 40 -- 150 keV\\
 & cts/s & cts/s\\
\hline
\hline 
LMC X-4 & 0.9(1) & 0.7(1)\\
Cen X-3  & 0.27(8) & --\\
4U 1700-377  & 1.31(8) & 0.50(6)\\
4U 1538-522 & -- & --\\
SMC X-1  & 0.48(16) & --\\
SAX J1802.7-2017  & 0.35(7) & --\\
XTE J1855-026  & -- & --\\
Vela X-1  & 1.02(7) & 0.15(6)\\
EXO 1722-363 & -- & --\\
OAO 1657-415 & 0.60(7) & 0.23(7)\\
\hline
\end{tabular}
\end{center}
\label{table:occultations_flux}
\end{table}
The average phase $\phi_{\rm (i,e)}$ (defined as the phase at which 99\% of the source flux is occulted at eclipse ingress or egress) and the transition 
phase width $\tau_{\rm (i,e)}$ were estimated by fitting independently the ISGRI light curve in the 17--40 keV and (whenever possible)  
in the 40--150 keV band with the function:  
\begin{equation}
F_{j} = F_{\rm cons}\, {\rm exp}\,\{ {\rm ln}(0.01)\,{\rm exp}\,[-(\phi_{j}-\phi_{\rm (i,e)})/\tau_{\rm (i,e)}]\}.
\label{eq:4}
\end{equation}
Here $j$ is the phase bin number and $F_{\rm cons}$ is the averaged count-rate of the source outside the eclipse determined prior to the fit. 
The 0.5--1.0 orbital phase was considered to fit the eclipse ingress and the 1.0--1.5 orbital phase was used for the egress. 
The duration of each X-ray eclipse for the ten sources was calculated as 
$\Delta\varphi = (1.0 -\phi_{\rm i})  + \phi_{\rm e}$ in the (17--40 keV) band, thus providing the semi-eclipse 
angles, $\theta_{\rm e}$, reported in Table \ref{obs:profile} (all uncertainties are given at $1 \sigma$ c.l.). 
We verified that the eclipse semi-angles measured in the 17--40 keV and 40--150 keV energy band are compatible to 
within the uncertainties. 
    
All the semi-eclipse angles we measured and reported in Table \ref{obs:profile} are a few degrees shorter than the values 
reported in the literature, the only two exceptions being 4U~1700-377 and LMC X-4. 
In the case of 4U~1700-377, our measured ingress phase is shorter than that measured by 
\citet{Rubin1996}, thus translating into a slightly longer eclipsing phase. For LMC~X-4 our determined semi-eclipse angle of 
$\theta_{\rm e}\approx16^{\circ}$ is around half the value $\sim27^{\circ}$ reported in the literature \citep{Li1978}. This is most 
likely due to the very few number of eclipses available in the past and the different energy band used to estimate $\theta_{\rm e}$ 
\citep{White1978b,Pietsch1985}. The usage of the long term observations available at present allow reducing the 
temporal variable distortions of the eclipse profile and provide a more reliable measurement of the semi-eclipse angle 
and eclipse duration.

\section{Discussion}
\subsection{Masses of the 10 Neutron Stars}
\label{sec:masses}

Determining the equation of state (EoS) of matter at densities comparable to those inside NSs is one of the most 
challenging problems of the modern physics and can only be addressed based on observations of astrophysical sources. 
Models proposed in the past years can be tested against observational results especially evaluating the 
maximum NS mass that each EoS model is able to sustain \citep[see e.g.,][]{Lattimer2001}. 
Very soft EoSs predict maximum NS masses in the 1.4--1.5 $M_{\odot}$ range (this occurs when the NS core is made by exotic 
matter as kaons, hyperons, and pions), whereas stiff EoSs can reach up to 2.4--2.5 $M_{\odot}$. More massive NSs can thus 
provide stronger constraints on the EoS models. As discussed by \citet{Rappaport1983}, eclipsing HMXBs hosting X-ray 
pulsars provide a means to measure the NS mass and thus put constraints on their EoS.  

In eclipsing HMXB the parameters needed to solve the equations that lead to the determination of the NS mass are: the system orbital period $P_{\rm orb}$, 
the projected semi-major axis $a_{\rm x}$sin$\,i$, the eccentricity $e$, the periastron angle $\omega$, and the duration of the eclipse  
expressed as the semi-eclipse angle $\theta_{\rm e}$ (see Sec. \ref{sec:semi-eclipse}). The semi-amplitude 
of the NS radial velocity can be inferred from the results of timing analysis providing  
$K_{\rm x} = 2\pi \,a_{\rm x}\mbox{sin}\,i\,/(P_{\rm orb}(1-e^2)^{1/2})$. The semi-amplitude of the radial 
velocity of the optical component, $K_{\rm opt}$, can be determined by optical and/or UV spectra of this star 
\citep[see Table \ref{table:nsmass}; we note that the values of $K_{\rm opt}$ determined through these techniques might be affected by uncertainties related to the companion star modelling\footnote{The strong stellar wind, which is responsible for the mass transfer, makes the task of generating a radial velocity curve of the donor star difficult. Another issue that can result in difficulties is also when the compact companion heats up one side of the donor star, thereby distorting spectral line profiles, i.e., X-ray heating take place.}; see,  e.g.,][]{Hammerschlag-Hensberge2003,Abubekerov2004,Koenigsberger2012}. 
The NS radial velocities, $K_{\rm x}$, reported in Table \ref{table:nsmass} are inferred using 
the values of $a_{\rm x}$sin$\,i$, $e$, and $P_{\rm orb}$ reported in Table~\ref{table1:sources} 
and \ref{table:results}.  
Two additional key parameters are required and can be estimated from theoretical arguments: $(i)$ the Roche 
lobe filling factor $\beta$ (i.e., the ratio of the supergiant's radius to that of its Roche lobe  
$\beta = R_{\rm opt}/R_{\rm L}$) and $(ii)$ the ratio of the spin period of the  supergiant to its 
orbital period $\Omega$. 

For sake of completeness, we review below the main equations to determine the NS masses \citep[see also][]{Rappaport1983}. 
The masses of the optical supergiant companion, $M_{\rm opt}$, and the neutron star, $M_{\rm x}$, can 
be written in terms of the mass functions: 
\begin{equation}
M_{\rm opt} = \frac{K_{\rm x}^{3}\,P_{\rm orb}\,(1-e^{2})^{3/2}}{2\pi\, \mbox{G} \,\mbox{sin}^{3}\,i}(1+q)^{2}
\end{equation}
and
\begin{equation}
M_{\rm x} = \frac{K_{\rm opt}^{3}\,P_{\rm orb}\,(1-e^{2})^{3/2}}{2\pi\, \mbox{G} \,\mbox{sin}^{3}\,i}(1+\frac{1}{q})^{2},
\end{equation}
where the mass ratio $q$ is defined as $q\equiv M_{\rm x}/M_{\rm opt} = K_{\rm opt} / K_{\rm x}$. Assuming a spherical 
companion star, the inclination angle of the system, $i$, is found to be related to the semi-eclipse angle through 
the geometric relation: 
\begin{equation}
\mbox{sin}\,i \approx \frac{\sqrt{1-\beta^{2}\, (R_{L}/a)^{2}}}{\mbox{cos}\,\theta_{\rm e}}. 
\end{equation}
We also made use here of the equation $\beta = R_{\rm opt}/R_{\rm L}$. The ratio between the Roche lobe radius and 
the separation of the two stars, $R_{\rm L}/a$, can be approximated as \citep{Joss1984}:
\begin{equation}
\frac{R_{\rm L}}{a}\approx A + B\,\mbox{log} \,q + C\,\mbox{log}^2 \,q,
\end{equation}
where: 
\begin{eqnarray}
A & = & 0.398 - 0.026\Omega^2 + 0.004\Omega^3\nonumber \\
B & = & -0.264 + 0.052\Omega^2 - 0.015\Omega^3 \nonumber \\
C & = & -0.023 - 0.005\Omega^2.
\end{eqnarray}
\begin{table*}
%\tiny
\caption{Best fit orbital profile parameters in units of the orbital phase. For each source we indicated with 
$\theta_{\rm e}$ the semi-eclipse angle measured from our ISGRI lightcurves (17--40 keV energy band), $\phi_{\rm i,e}$ is the 
defined as the phase at which 99\% of the source flux is occulted at eclipse ingress or egress, and $\tau_{\rm i,e}$ the corresponding transition 
phase width. The values of the 
semi-eclipse angles reported in the last column, $\theta_{\rm e,old}$, are from the literature. We indicated in 
brackets the uncertainties at $1\sigma$ c.l. on the last digits of each reported value.}
\begin{center}
\begin{tabular}{lllllll}
\hline
Source & $\phi_{\rm i}$  & $\tau_{\rm i}$ &  $\phi_{\rm e}$  &$\tau_{\rm e}$ & $\theta_{\rm e}$& $\theta_{\rm e,old}$\\
&&&&&(deg)&(deg)\\
\hline
\hline
LMC X-4 & 0.951(4) &  0.020(1) & 0.039(3) & 0.012(2) & 15.8(8) &27(2)$^{a}$\\
Cen X-3 &  0.922(1)& 0.019(1)&0.077(1)  & 0.0187(6)   &27.9(3) & 33(1)$^{b}$\\
4U 1700-377 & 0.912(4) & 0.0185(5) & 0.0894(4)& 0.0126(2)   & 32(1) & 29(2)$^{c}$\\
4U 1538-522 & 0.936(4) & 0.018(2)  &0.052(4)& 0.0147(5)  &21(1) & 29(2)$^{d}$\\
SMC X-1 & 0.939(9) & 0.0139(4) & 0.066(5) & 0.007(2)  &23(2) & 28(2)$^{a}$\\
SAX J1802.7-2017 &0.94(1)&0.038(9)   & 0.11(1)  &0.012(7)  & 31(2) & 35(5)$^{e}$\\
XTE J1855-026 &  0.920(5)& 0.0129(3) & 0.097(6)& 0.016(4) &32(1)& 42(6)$^{f}$\\
Vela X-1 & 0.9202(4) & 0.0209(1)  & 0.0899(2) & 0.0084(2) &30.5(1) & 33(3)$^{g}$\\
EXO 1722-363 & 0.94(2) & 0.013(8)  & 0.088(4) & 0.004(2) &26(4) &32(2)$^{h}$\\
OAO 1657-415 & 0.933(1) &0.0184(6)  & 0.049(2)& 0.0352(1)  &20.9(4) & 30(1)$^{i}$\\
\hline
\end{tabular}
\tablefoot{$^{a}$\citet[][and references therein]{vanderMeer2007}, $^{b}$\citet{Clark1988}, 
$^{c}$\citet{Rubin1996}, $^{d}$\citet{vanKerkwijk1995b}, $^{e}$\citet{Hill2005}, 
$^{f}$\citet{Corbet2002}, $^{g}$\citet{vanKerkwijk1995a}, 
$^{h}$\citet{Corbet2005}, $^{i}$\citet{Chakrabarty1993}.}
\label{obs:profile}
\end{center}
\end{table*}

For HMXBs with circular orbits the ratio $R_{\rm L}/a$ is constant. When the orbit is eccentric the Roche Lobe filling factor, 
$\beta$, is defined at the periastron, $R_{\rm L}$, and the separation between the centers of masses of the two 
stars varies with the orbital phase. n these cases, the separation between the centers of the two stars at mid-eclipse time is given by 
$a'=a(1-e^2)/(1+e\,cos\,\omega)$, where $\omega$ is the argument of periastron reported in Table \ref{table1:sources}.
In our calculation the Roche lobe radius is estimated at the eclipse phase. The approximated Roche lobe 
radius, $R_{\rm L}$, is determined with an accuracy of about 2\% over the ranges of $0\leqslant \Omega \leqslant2$ and 
$0.02 \leqslant q \leqslant 1$ \citep{Joss1984}. 
Once the full set of input parameters above are given, the values of $i$, $M_{\rm x}$, $M_{\rm opt}$, $a$, $R_{\rm L}$, 
and $R_{\rm opt}$ can be determined at $1 \sigma$ c.l. by means of Monte-Carlo (MC) simulations. Uncertainties on all parameters 
are evaluated in the simulations by assuming that their values follow a Gaussian distribution.  
We used a linear distribution for the Roche lobe filling factor, $\beta$, spanning the range 0.9--1.0. 
Note that for all the HMXBs considered here it is known from their optical lightcurves that the supergiant stars hosted in these systems  
are nearly filling their Roche lobes \citep[see e.g., ][and references therein] {Tjemkes1986}.

Following the technique described above, we first computed for the four sources with known $\Omega$ (LMC X-4, Cen X-3, SMC X-1, and Vela X-1) 
the remaining output parameters reported in Table \ref{table:nsmass}. In order to verify the reliability of this method, 
we compared the observational reported projected stellar rotation velocity, $v_{\rm rot} \mbox{sin} \,i$,  measurements 
with our calculated value using results in Table \ref{table:nsmass} and the following equation:
\begin{equation}
\Omega = 0.02\biggl( \frac{v_{\rm rot} \mbox{sin} \,i}{\mbox{km s}^{-1}}\biggr)\biggl( \frac{P_{\rm orb}}{\mbox{days}}\biggr)
\biggl( \frac{R_{\rm opt}}{R_{\odot}}\biggr)^{-1}\frac{(1-e)^{3/2}}{(1+e)^{1/2}}.
\label{Eq:Vrot}
\end{equation}
In Table \ref{table:comparison} we compare the calculated and measured $v_{\rm rot} \mbox{sin} \,i$ values. 
These values agree well within the uncertainties. 
Based on these results, we assumed for the sources 4U 1538-522, SAX J1802.7-2017, XTE J1855-026, EXO 1722-363, 
and OAO 1657-415 a mean value of $\Omega=0.91(20)$. As these systems are not circularized, $\Omega$ 
is unlikely to be too close to unity. If tides are efficient, then we might expect that the rotation 
of the donor is synchronized at periastron, where the tides are most effective \citep{HUT1981}. 
In this case, the star's angular velocity is larger than the orbital velocity at mid-eclipse, 
thus leading to $\Omega \gtrsim 1$. For this reason we increased the uncertainty on $\Omega$ 
to include values slightly larger than unity. 

4U 1700-377 was considered separately, as its projected semi-major axis $a_{\rm x}$sin$\,i$ is unknown and thus the corresponding
$K_{\rm x}$ cannot be determined as we have done for all other sources. For this source, we first estimated $\Omega$ from
Eq. \ref{Eq:Vrot} by using the observationally measured value of $v_{\rm rot} \mbox{sin} \,i$, $P_{\rm orb}$, and
the companion star radius $R_{\rm opt}=21.9(1.9) R_{\odot}$
\citep[see Table \ref{table:nsmass} and][]{Clark2002}. We thus performed several MC simulations with a variable $K_{\rm x}$, until the $R_{\rm opt}$ obtained from one of the simulations was compatible (to within the uncertainties) with the observational value. The outputs of this simulation were then used to fill all other relevant parameters for the source 4U 1700-377 in Table~\ref{table:nsmass}.

\begin{figure} 
\psfig{figure=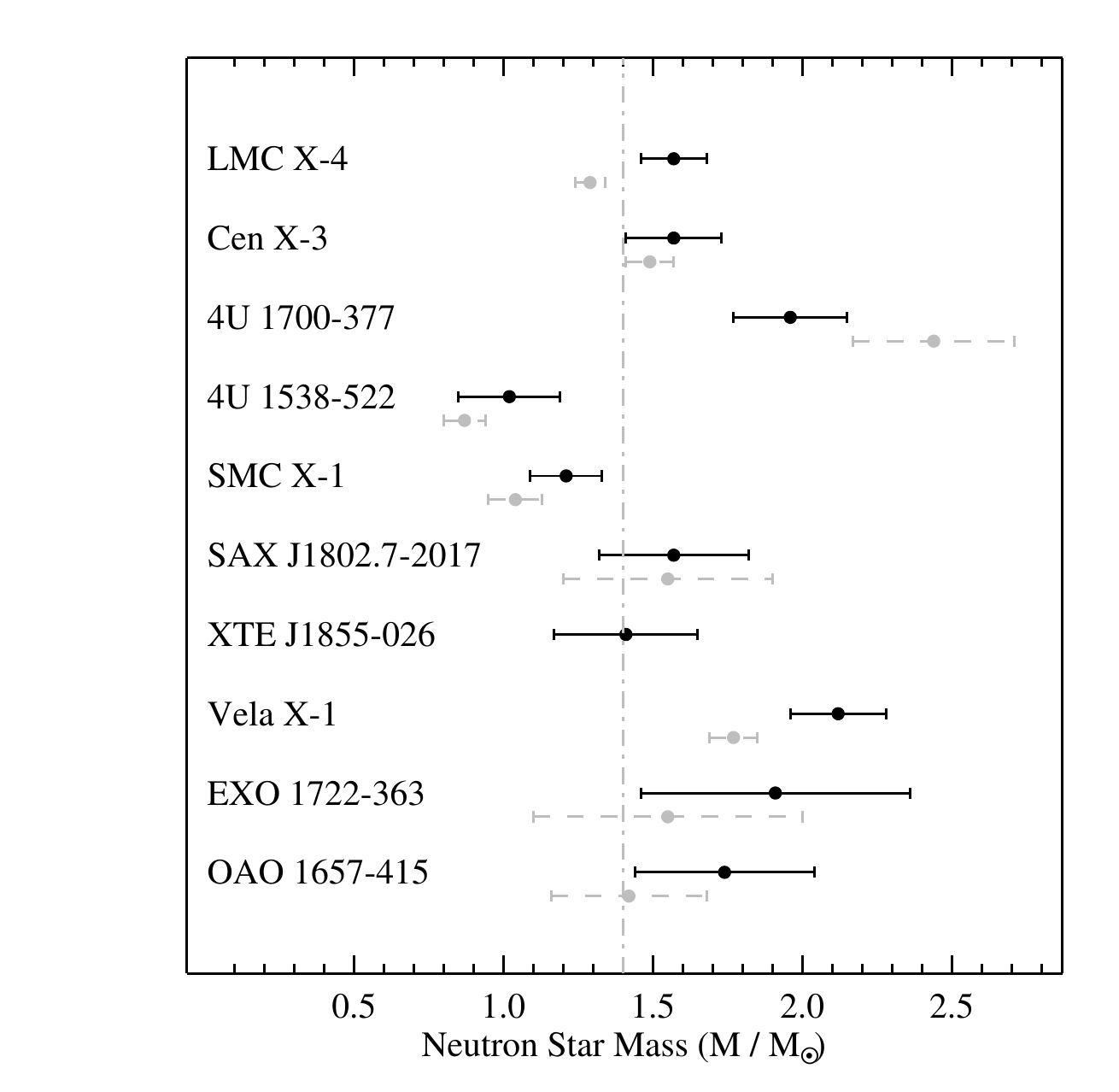,width=9.0cm}
\caption[]{The masses of the ten eclipsing HMXBs. The neutron star masses determined in this work are shown with solid lines. 
Values from the literature are represented with dashed lines. The error bars correspond to uncertainties at $1 \sigma$ c.l.
The dashed vertical line indicate the canonical neutron star mass of 1.4 M$_{\odot}$.}
 \label{fig:Mass}
\end{figure}

Our findings on all neutron star masses for the considered HMXBs are summarized in Fig. \ref{fig:Mass}, together
with previously published values. The neutron star masses for LMC X-4, Cen X-3, 4U 1538-52, SMC\,X-1, and Vela X-1 are
from \citet{Rawls2011}, while the mass of the neutron star hosted in 4U 1700-377 is derived from
\citet{Clark2002}. The corresponding values for SAX J1802.7-2017, EXO 1722-363, and OAO 1657-415 are obtained
from \citet{Mason2011,Mason2010,Mason2012}. We note that our estimated uncertainties on the neutron star masses
are somewhat more conservative than those reported by \citet{Rawls2011} and slightly smaller than those
obtained by \citet{Clark2002} and \citet{Mason2011,Mason2010,Mason2012}. The reason behind these differences is that
these authors derived their uncertainties either from analytical calculations or through numerical simulations
including a particularly refined treatment for the Roche-Lobe size of the supergiant star. In the present work,
all uncertainties are, instead, derived directly from the MC simulations and account for the most recently
updated system parameters obtained from multi-wavelength observations. In particular, we reported for all sources 
in Table \ref{obs:profile} the most accurately measured values of the semi-eclipse angles, which play a crucial role in the 
MC simulations. We note that in all cases our measured semi-eclipse angle is smaller than reported values  
in the literature (see Table \ref{obs:profile}), and thus the NS masses we estimated are generally larger. This seems reasonable 
especially for 4U\,1538-522 and SMC\,X-1, as the NS masses estimated before where $\lesssim$1\,M$_{\odot}$. 
The only exception to this trend is 4U\,1700-377, for which our estimated semi-eclipse angle is comparable to 
that obtained by \citet{Rubin1996}.   
    
\begin{table*}
\tiny
\caption{Input and output parameters for the MC simulations used to finally estimate the mass of the NSs hosted in the 10 HMXBs considered 
in this work. We indicated in brackets the uncertainties at $1\sigma$ c.l. on the last digits of each reported value.}
\begin{center}
\begin{tabular}{llllllllllll}
\hline
  & & Input  &&&&&&Output &&&
 \\ \cline{2-5}  \cline{7-12}
Source & $K_{\rm x}$ & $K_{\rm opt}$ & $q$& $\Omega$&    &$i$ & $M_{\rm ns}$  & $M_{\rm opt}$ &  $R_{\rm opt}$ & $R_{\rm L}/a$ & $a$
\\ \cline{2-5}  \cline{7-12}
 & km s$^{-1}$ & km s$^{-1}$ & && & deg & $M_{\odot}$ & $M_{\odot}$  & $R_{\odot}$ &  & $R_{\odot}$\\
\hline
\hline
LMC X-4 & 407.8(3)& 35(2)$^{a}$& 0.086(4)& 0.97(13)$^{a}$& & 59.3(9)& 1.57(11) &18(1)& 7.4(4)  &0.59(1) & 14.2(2)$^{l}$\\
Cen X-3 & 414.317(9) & 28(2)$^{a}$ & 0.066(5)& 0.75(13)$^{a}$ &&65(1) &  1.57(16)&24(1) &11.4(7)  &0.63(1) &20.2(4) \\
4U 1700-377$^{*}$ & 435(10) & 19(1)$^{b}$&0.043(2)& 0.47(4)$^{cc}$&  &  62(1)& 1.96(19)& 46(5)&22(2) & 0.694(6)& 35(1)\\
4U 1538-522 & 316(10)& 20(3)$^{c}$& 0.06(1)&0.91(20)$^{m}$& & 67(1) & 1.02(17) &16(2)& 13(1) & 0.53(3)&22(1)$^{l}$\\
SMC X-1 & 299.631(5)& 20(1)$^{a}$&0.067(4) &0.91(20)$^{a}$ & &62(2) &1.21(12) &18(2)& 15(1)  & 0.61(2)& 27.9(7)$^{l}$\\
SAX J1802.7-2017 & 324(5)& 24(3)$^{d}$& 0.07(1) & 0.91(20)$^{o}$ &&72(2) & 1.57(25)& 22(2) & 18(1)& 0.61(2)& 33(1)\\
XTE J18f55-026 & 289(5) & 20(3)$^{n}$ & 0.07(1) & 0.91(20)$^{o}$ && 71(2) & 1.41(24)& 21(2) & 22(2) & 0.63(3) & 40(1)$^{l}$\\
Vela X-1 & 278.1(3) & 23(2)$^{e}$& 0.081(5)& 0.67(4)$^{h,i}$ && 72.8(4)&  2.12(16) & 26(1)& 29(1)  &  0.595(6)& 59.6(7)$^{l}$\\
EXO 1722-363 & 226(2)& 25(5)$^{f}$ & 0.11(2) & 0.91(20)$^{o}$ && 68(2) & 1.91(45) & 18(2) & 26(2) & 0.58(3) & 52(2)\\
OAO 1657-415 & 222.60(6)& 21(4)$^{g}$&0.09(1) & 0.91(20)$^{o}$ && 67.9(9) & 1.74(30) & 17.5(8) & 25(2) & 0.52(2) & 53.1(8)$^{l}$\\
\hline
\end{tabular}
\tablefoot{$^{a}$\citet{vanderMeer2007},$^{b}$\citet{Hammerschlag-Hensberge2003}; $^{c}$\citet[][see reference therein]{vanKerkwijk1995b}, 
$^{cc}$$v_{\rm rot} \mbox{sin} \,i = 150(10)$ km/s (Clark et al, 2002); $^{d}$\citet{Mason2011}; $^{e}$\citet{Quaintrell2003}; 
$^{f}$\citet{Mason2010}, $^{g}$\citet{Mason2012}; $^{h}$\citet{Howarth1997}; $^{i}$\citet{Zuiderwijk1995}; 
$^{l}$ evaluated at periastron; $^{m}$we used the same $\Omega$ as for SMC X-1 since the values of  
$v_{\rm rot} \mbox{sin} \,i$ for these systems are compatible to within the uncertainties \citep[see also][]{Rawls2011}; 
$^{n}$we assumed a value of $K_{\rm opt}$ derived from the average of the other systems in this table with properties close to XTE J1855-026; $^{o}$we assume a conservative $\Omega$ value in order to be close to the Roche lobe radius; $^{*}$ we treat 4U 1700-377 differently, since the project semi-major axis, $a_{\rm x}$sin$\,i$, is unknown; see Sec. \ref{sec:masses}.}
\label{table:nsmass}
\end{center}
\end{table*}

\begin{table}
\tiny
\caption{Comparison between the calculated and measured values of $v_{\rm rot} \mbox{sin} \,i$.}
\begin{center}
\begin{tabular}{lcc}
\hline
Source & $v_{\rm rot} \mbox{sin} \,i$& $v_{\rm rot} \mbox{sin} \,i^{*}$\\   
 & km s$^{-1}$ & km s$^{-1}$\\
& & $(\beta = 0.9-1.0)$\\
\hline
\hline
LMC X-4 &  240(25)$^{a}$		&257.9(7) \\
Cen X-3 &  200(40)$^{a}$		&204.8(8)\\
4U 1538-522 &180(30)$^{b}$	&224.9(7)\\
SMC X-1 & 170(30)$^{a}$ 	&178.0(8)\\
SMC X-1 & 172(1.5)$^{c}$ 	&178.0(8)\\
Vela X-1 & 116(6)$^{d}$	&130.3(2)\\
\hline
\end{tabular}
\tablefoot{$^{a}$\citet{vanderMeer2007}, $^{b}$\citet[][see reference therein]{vanKerkwijk1995b}, $^{c}$\citet{Reynolds1993}, 
$^{d}$\citet{Zuiderwijk1995}. $^{*}$This work, using Eq. \ref{Eq:Vrot} and our results reported in 
Table  \ref{table:nsmass} and Table \ref{table:results}.}
\label{table:comparison}
\end{center}
\end{table}

\subsection{Orbital period change}
\label{sec:orbital_decay}

Five out of the 10 HMXBs analyzed in the present work showed evidence for a
significant orbital period decay, i.e., LMC\,X-4, Cen\,X-3, 4U\,1700-377, SMC\,X-1, and OAO\,1657-415 (our values can be compared with previous measurements published in the literature by using the references reported in the Tables in Appendix~\ref{table:ephemerides}).
For all these sources, a number of different mechanisms to explain the orbital period decay has been invoked \citep[see e.g.,][]{Kelley1983,Vanderklis1983,vanderKlis1984,Levine1993,Rubin1996,SafiHarb1996,Levine2000,Jenke2011}.
The hypothesis that the period change is due entirely to angular momentum loss in the stellar wind,
i.e. mass transfer with no interaction, can be ruled out. More realistic models proposed that the orbital decay is mostly driven by the tidal interaction and the rapid mass transfer between the two objects. The latter takes place due to the fast winds that characterize most of the donor stars in HMXBs. In these models, the asynchronism between the orbit and the rotation of the donor star is maintained by the evolutionary expansion of the donor star \citep[see e.g.,][and references therein]{Lecar1976,SafiHarb1996,Levine2000,vanderKlis1984}. If the donor star is  rotating with an angular velocity close to synchronization, the matter ejected leaves the system and carries an extra momentum due to this rotation. This loss of rotational momentum affects the orbital period only if there is a coupling between the rotation of the donor star and the orbit of the neutron star. Tidal interactions could synchronize the orbit with the rotation of the donor star and allow the transfer of rotational momentum of the primary star to the orbital momentum. Tidal effects on the radial velocity curve of the donor star in Vela X-1 have been investigated deeply, but not final conclusions could be drown due to the limited quality of the data \citep{Koenigsberger2012}. Tidal interactions in close binary systems has been investigated in detail also theoretically by using different approaches, e.g., assuming the rotation of the donor star is synchronized or pseudosynchronized\footnote{Pseudosynchronized systems are binaries in which the donor star is synchronized only at the periastron, where the tides are more effective \citep{HUT1981}.} with the orbital motion of the compact star
\citep[see e.g.,][]{Moreno2011,HUT1981,Zahn1977,Lecar1976}. 
A number of other works argued that the combined effects of the evolutionary growth of the moment of inertia (expansion of the donor star) and mass loss by stellar wind from a donor star can explain the orbital decay in HMXBs \citep{bagot1996,Levine1993}. In these models, it is considered that, as the companion star evolves, it expands and increases its moment of inertia, leading to a slow down of the star rotation. Tidal torques would then transfer orbital momentum to synchronize the binary system, finally causing an orbital decay \citep[see e.g., ][and references therein]{Levine1993}.  As the lifetime of a supergiant star does not exceed a few 10$^6$~yr, we expect that most of the systems considered in this work are nearly synchronized but not fully circularized yet. This is in agreement with the non-negligible values of the eccentricity measured in some of the eclipsing HMXBs (see Table~\ref{table1:sources}).

The updated and newly reported estimates of the orbital periods decay of the ten HMXBs considered in the present work can be used as inputs in future works on these topics. We expect that this might help discriminating among the different mechanisms proposed to explain the orbital period decay in HMXBs and to estimate their orbital circularization and synchronization timescales, as well as to discuss the Darwin instability \citep[see e.g.,][]{Lai1994} for these systems. Additional calculations on all these effects are beyond the scope of the present paper.

In order to help future works to evaluate numerically different theoretical models, we finally report for reference also the stellar wind velocities and mass loss rates of all sources considered in this paper in Table~\ref{table:puls}.
These have been calculated by assuming the spectral types reported in Table~\ref{table1:sources}. In Table~\ref{table:puls} we indicate both the terminal wind velocities inferred from observations in the literature and those calculated theoretically. For the latter we followed \citet{Martins2005}
and \citet{Kudritzki2000} for O-stars, and \citet{Markova2008}, \citet{Crowther2006}, \citet{Lefever2007} for B-stars.
In all cases, the nominal mass loss rates have been reduced by a factor of $\sim3$ in order to account for clumping and uncertainties
on the terminal wind velocities are assumed to be at the 30\% \citep{Markova2008,Repolust2004}. As all data in Table~\ref{table:puls} have been calculated assuming the wind is emitted by an isolated massive star, it is possible that in a number of systems the disturbance of the neutron star lead to substantially different wind properties. Detailed numerical simulations have showed that this could be the case especially for short orbital period systems
with high X-ray luminosities \citep{Watanabe2006,Manousakis2014}. Unfortunately, direct measurements of the wind properties for the systems
considered here are challenged by the relatively high absorption local to the sources and their large distance \citep{Chaty2013}.

\begin{table}
\tiny
\begin{center}
\caption{Properties of the stellar winds in the 10 HMXBs considered in this paper as inferred from observations and 
theoretical calculations. }
\label{table:puls}
\begin{tabular}{lccc}
\hline
Source & $v_{\infty}$ & $v_{\infty}$ & $\dot{M}$ \\   
 & km~s$^{-1}$ & km~s$^{-1}$& M$_{\odot}$~yr$^{-1}$ \\
  & Observed & Theoretical$^{*}$ & Theoretical$^{*}$\\
\hline
\hline
LMC X-4 & 1350(35)$^1$& 1950(600) & $2.4\times10^{-7}$\\
Cen X-3 &  $\sim$500$^2$ & 2050(600) & $5.3\times10^{-7}$ \\
4U 1700-377 & 1700(100)$^3$ &1850(550)& $>2.1\times10^{-6}$ \\
4U 1538-522 & --$^4$ &1000(300)& $8.3\times10^{-7}$  \\
SMC X-1 & --$^4$ &870(260) & $1.5\times10^{-6}$\\
SAX J1802.7-2017 & --$^4$ & 680(200)& $6.3\times10^{-7}$\\
XTE J1855-026 & --$^4$ & 620(190)&$<(0.2-1.1)\times10^{-5}$ \\
Vela X-1 & 600(100)$^3$ & 640(190) & $<(1.0-5.3)\times10^{-6}$ \\
EXO J1722-363 & --$^4$ & 650(200) & $9.0\times10^{-7}$ \\
OAO 1657-415 & $\sim$250$^5$  &200(60) & $<(1.1-5.6)\times10^{-7}$ \\
\hline
\end{tabular}
\tablefoot{References: {$^1$} \citet{boroson99}; {$^2$} \citet{wojdowski03}; {$^3$} \citet{vanloon2001}; {$^4$} No consolidated measurement 
available in the literature - at the best of our knowledge; {$^5$} \citet{Mason2012}.}
\end{center}
\end{table}

\subsection{Apsidal motion}
\label{sec:apsidal2}

Beside asteroseismology \citep[see, e.g.,][and references therein]{Dupret2011}, the measurement of the apsidal advance 
in eccentric binary systems offer an alternative possibility
to investigate the internal structure of stars compared to theoretical calculations
and numerical modelling. The rate at which the longitude of periastron of an eccentric orbit advances can
be related to the absidal motion constant, which depends on the model of the mass distribution inside the
star \citep{Kopal1978,Batten1973}. X-ray pulsars in HMXBs have been far considered the most promising laboratories
to measure accurately the apsidal advance. The neutron star in these systems can be well approximated as a point
source with a negligible absidal motion constant compared to that of the massive companion.
The latter is thus dominating the apsidal advance \citep{Rappaport1980}.

The long term analysis carried out in this paper, allowed us to provide the most accurate apsidal advance measurement
for Vela\,X-1 and 4U\,1538-522. (note, however, that for the former source the measurement is only marginally significant 
at 1.5$\sigma$).

In the case of Vela\,X-1, the theoretical estimated apsidal motion is $\dot{\omega}$$\sim0.4{\degr}$ yr$^{-1}$, assuming
a companion star surface temperature of $25000$~K \citep[as predicted accordingly to its spectral type; see, e.g.,][]{Avni1975,Conti1978}.
The measurement reported in Table~\ref{table:apsidal} is thus fully consistent with the theoretical expected value
\citep[see][and references therein]{Deeter1987b}. To the best of our knowledge, this is the first time that such
match is reported, given the reduced uncertainty on the measured value of $\dot{\omega}$ (even though this is still only marginally 
significant). 

In the case of 4U\,1538-522, no firm measurement of the apsidal motion was reported in the literature.
We notice that the tentative value provided by \citet{Mukherjee2006} was estimated by comparing only two measurements
of $\omega$ collected in 1997 and 2003 and thus is not comparable to the refined estimate reported in the present work.
Our measurement thus provides the first consolidated estimate of $\dot{\omega}$ for 4U\,1538-522.

\subsection{Eclipse asymmetry}
\label{sec:eclipse}

The updated ephemeris for all the 10 HMXBs in Table~\ref{table:results} that 
we obtained in this work allowed us to fold with an unprecedented accuracy the 
long term ASM and ISGRI lightcurves of these objects in different energy bands. 
The results are shown in Fig.~\ref{fig:lightcures} and \ref{fig:lightcures2}.  
The available broad-band energy coverage permitted to study in detail
the averaged profile of the eclipse ingress at egress for each object.
We note that in the case of SAX\,J1802.7-2017, XTE\,J1855-026,
and EXO\,1722-363, this study could not be easily carried out at low energies due to
the strong absorption affecting the X-ray emission from these sources
\citep[this is expected according to their classification as ``highly obscured
HMXBs''; see, e.g.][for a recent review]{Chaty2013}. In most of the other sources we found peculiar asymmetries that are more pronounced at the lower
energies and slightly more enhanced in objects characterized by a non-negligible eccentricity.
Remarkable asymmetries are recorded for the eclipse ingresses and egresses of 4U\,1700-377 and
Vela\,X-1. Evident eclipse asymmetries are also displayed by 4U\,1538-522 and OAO\,1657-415.

In most of the sources considered in this work, single eclipse profiles have been studied 
in the past by using focusing X-ray instruments capable of high resolution spectroscopy \citep[e.g., XMM-Newton and Chandra; see]
[and references therein]{Watanabe2006}. These observations proved to be particularly powerful to analyze how the X-ray emission from 
HMXBs is affected by the presence of inhomogeneities in the stellar winds \citep{sako2003}. Such inhomogeneities 
\citep[``clumps'', see e.g.][for a recent review]{Puls2008} are transported away from the star with typical  
velocities of a smooth wind and thus can induce changes in the mass accretion rate (and X-ray luminosity) on timescales as short as  
100-1000~s. The feedback of X-ray radiation is also known to affect the structure of the wind on similar time scales, as high energy photons 
ionize the metal ions in the wind and dramatically reduce the wind acceleration mechanism (in turns affecting the velocity and 
density profile of outflows from the massive stars).   

The eclipse profiles we discussed above provide, instead, important information on the interaction between 
the compact object and the stellar wind on much longer time scales ($\sim$yrs). The influence of wind inhomogeneities 
is in all cases averaged away by the long-term integration, and only the effect of X-ray irradiation on the wind 
averaged on hundreds of orbital revolutions can be observed from these curves. 
Under these assumption, we argue that the eclipse asymmetries are most likely due to the presence of ``accretion wakes''. 
These structures are commonly observed in HMXBs and form as 
a consequence of the compact object intense gravity and conspicuous X-ray emission (see, e.g., Fig. 5 in \citet{Blondin1990} and Fig.1-5 in \citet{Manousakis2012}).

As the accretion wake usually trails the neutron star during its orbital revolution around the companion and can lead to an 
increase of the wind density around the compact object by a factor of $\sim$10-100 \citep{Manousakis2014}, it is expected that 
absorption column density in the direction of the X-ray source progressively increases before the occurrence of the X-ray eclipse. 
During the egress from the eclipse, the accretion wake is located beyond the compact object along the line of sight to the observer 
and thus do not lead to any apparent increase of the local absorption column density. 

An interesting possibility is also that this effect is enhanced in systems endowed with a non-negligible orbital eccentricity 
due to the different degree of ionization of the stellar wind material when the neutron star 
approaches or recedes from the companion. It is well known that in case of eccentric orbits, the absolute value of the 
vectorial sum of the neutron star orbital velocity and the stellar wind velocity reaches a minimum while the compact object 
approaches the companion, and then increases toward the upper conjunction 
\citep[see, e.g.,][and references therein]{Ducci2010}. In a wind accreting system, the amount of material that a neutron 
star can capture from the companion at any time is roughly comparable to $\pi$$R_{acc}^2$, where 
\begin{equation}
R_{\rm acc}\simeq\frac{2G M_{\rm NS}}{v_{\rm rel}^2}
\label{eq:raccr}
\end{equation}
and $v_{\rm rel}$ is the relative velocity mentioned above. It is thus expected that a larger X-ray luminosity is released 
at lower values of $v_{\rm rel}$ due to the enhanced accretion rate \citep[see, e.g.,][and references therein]{Bozzo2008}. 
As the ionization of the wind is proportional to the X-ray flux and acts to substantially inhibit the wind acceleration mechanism, 
a larger density would also be expected around a bright X-ray emitting neutron star approaching the companion. Furthermore, the 
X-ray flux could be evaporating additional material from the massive star around this orbital phase, possibly enhancing the absorption 
column density in the direction of the source during the eclipse ingress \citep{Blondin1995}. We note that in short orbital period systems, 
the formation of (at least) temporary accretion disks when the neutron star is at the closest distance from the companion could 
also provide enhanced accretion rates and thus positively contribute to increase the mass density around the neutron star \citep[e.g.,][]{Ducci2010}.  
A quantitative estimate of all these effects would require more detailed calculations combining also the effect of the neutron star 
spin and magnetic field. 

\section{Conclusions} 
\label{sec:conclusions} 

In this work, we performed orbital mid-eclipse time measurements on the ten
known eclipsing HMXBs using publicly available pre-processed \rxte/ASM and \I/ISGRI data.
For each source, we determined several new orbital mid-eclipse
times and used them together with all previously published value (to the best of our knowledge) to
obtain the most updated and homogeneous set of ephemerides to date.
The latter allowed us in particular to update the orbital periods of all the 10 binaries, as well as the
decay of these periods. In five sources (LMC\,X-4, Cen\,X-3, 4U\,1700-377, SMC\,X-1, and OAO\,1657-415) we measured a significant orbital period decay, with values in the range
$1.0-3.5\times10^{-6}$ yr$^{-1}$; in all other case no significant decay was revealed even though more than 30 yrs
of X-ray observations were used. For the eccentric systems Vela\,X-1 and 4U~1538-522 we accurately determined for
the first time a significant apsidal advance, comparing it with previously expected values.

Using our best estimated mid-eclipse epoch, orbital period, and period derivative for each source, we
folded the lightcurves of the ten binaries obtained from the long-term monitoring carried out with the ASM on-board RXTE
(1.3-12 keV) and ISGRI on-board INTEGRAL (17-150 keV). We discussed the asymmetric profiles of the eclipse ingresses
and egresses, comparing them with theoretical expectations from simulations of accreting wind-fed systems.

Finally, we used all above mentioned results within the MC simulations to consistently and systematically evaluate the masses of the
NS hosted in the 10 eclipsing HMXBs.The results reported in this paper constitute an important database for population and evolutionary studies of 
HMXBs, as well as theoretical modelling of long-term accretion in wind-fed X-ray binaries.

\begin{acknowledgements} 
MF acknowledge ISSI (Switzerland) for financial support during the visit   
of R. Vuillez in Bern, and thank F. Mattana, R. Vuillez, J. A. Zurita-Heras, and 
A. Goldwurm for their contributions on an earlier version of the manuscript. 
AL acknowledge support by the grant RFBR 12-02-01265. 
We acknowledge support from ISSI through funding for
the International Team meeting on Unified View of Stellar
Winds in Massive X-ray Binaries (ID 253).
\end{acknowledgements} 

\begin{appendix} 

\section{Ephemeris tables}
\label{table:ephemerides}
%MJD=JD - 2400000.5
\begin{table}[h]
\tiny
\begin{centering}
\begin{tabular}{lll}
\multicolumn{3}{c}{LMC X-4}\\ 
\hline
Epoch Time&  Satellite & Reference\\
 (MJD) & & \\
\hline
\hline 
42829.494(19) & {\em SAS-3} & \citet{Kelley1983} \\
45651.917(15)  & {\em EXOSAT} & \citet{Dennerl1991}\\
45656.1453(8) & {\em EXOSAT}  & \citet{Dennerl1991}\\
46447.668(11)  & {\em EXOSAT} &  \citet{Dennerl1991}\\
46481.467(3)  & {\em EXOSAT} &  \citet{Dennerl1991}\\
47229.3313(4)  & {\em Ginga} & \citet{ Woo1996}\\
47741.9904(2)  & {\em Ginga} & \citet{Levine1991}\\
48558.8598(13) & {\em ROSAT} & \citet{Woo1996}\\
51478.454(8) & \rxte/ASM & Present Work \\
51110.86579(20)  & \rxte & \citet{Levine2000}\\
52648.813(14) & \I & Present Work \\
53013.590(14) & \I & Present Work \\
53016.40(2)  & \I & Present Work \\
54262.825(8) & \rxte/ASM & Present Work \\
\hline
\end{tabular}
\end{centering}
\end{table}

\begin{table}
\tiny
\begin{centering}
\begin{tabular}{lll}
\hline
\multicolumn{3}{c}{Cen X-3}\\ 
\hline
Epoch Time&  Satellite & Reference\\
$T_{\pi/2}$ &&\\
\hline 
\hline
40958.34643(45) & {\em Uhuru} & \citet{Fabbiano1977}\\
41077.31497(15) &  {\em Uhuru} & \citet{Fabbiano1977}\\
41131.58181(29) &  {\em Uhuru} & \citet{Fabbiano1977}\\
41148.28051(16) & {\em Uhuru} & \citet{Fabbiano1977}\\
41304.81533(14) &  {\em Uhuru} & \citet{Fabbiano1977}\\
41528.1401(3) &  {\em Uhuru} & \citet{Fabbiano1977}\\
41551.09798(17) &  {\em Uhuru} & \citet{Fabbiano1977}\\
41569.88199(11) &  {\em Uhuru} & \citet{Fabbiano1977}\\
41574.05610(13) &  {\em Uhuru} & \citet{Fabbiano1977}\\
41576.1433(1) &  {\em Uhuru} & \citet{Fabbiano1977}\\
41578.23037(7) &  {\em Uhuru} & \citet{Fabbiano1977}\\
41580.31722(9) &  {\em Uhuru} & \citet{Fabbiano1977}\\
41584.49193(10) &  {\em Uhuru} & \citet{Fabbiano1977}\\
41590.75328(15) &  {\em Uhuru} & \citet{Fabbiano1977}\\
41592.84025(15) &  {\em Uhuru} & \citet{Fabbiano1977}\\
41599.10212(15) &  {\em Uhuru} & \citet{Fabbiano1977}\\
41601.18930(14) &  {\em Uhuru} & \citet{Fabbiano1977}\\
41603.27671(21) &  {\em Uhuru} & \citet{Fabbiano1977}\\
42438.128(3) & {\em Ariel-V} & \citet{Tuohy1976} \\
42786.6755(7)  & {\em Cos-B} & \citet{vanderKlis1980}\\
43112.26642(40)  & {\em SAS-3} & \citet{Kelley1983}\\
43700.83275(43)  & {\em HEAO-1} & \citet{Howe1983}\\
43869.88910(2)  & {\em SAS-3} & \citet{Kelley1983}\\
44685.94760(5)  & {\em Hakucho} & \citet{Murakami1983}\\
45049.1025(1)  & {\em Hakucho} & \citet{Nagase1984PASJ}\\
45428.95421(5)  & {\em Tenma} & \citet{Nagase1984PASJ}\\
47607.8688(8)  & {\em Ginga} & \citet{Nagase1992}\\
50506.788423(7)  & {\em RXTE} & \citet{Raichur2010}\\
50782.279(8) & \rxte/ASM & Present Work \\
52180.589(8) & \rxte/ASM & Present Work \\
53136.455(14)  & {\em INTEGRAL} & Present Work\\
53574.711(8) & \rxte/ASM & Present Work \\
54144.471(14)  & {\em INTEGRAL} & Present Work\\
54966.745(14) & \rxte/ASM & Present Work \\
\hline
\end{tabular}
\end{centering}
\end{table}

\begin{table}[h]
\tiny
\begin{centering}
\begin{tabular}{lll}
\multicolumn{3}{c}{4U 1700-377}\\ 
\hline
Epoch Time&  Satellite & Reference\\
$T_{\pi/2}$ &&\\
\hline 
\hline
41452.64(1)  & {\em Uhuru} & \citet{Jones1973}\\
42609.25(1) & {\em Copernicus} & \citet{Branduardi1978}\\
42612.646(10)  & {\em Copernicus} & \citet{Branduardi1978}\\
43001.604(10) & {\em Copernicus}  & \citet{Branduardi1978}\\
43005.00(1)  & {\em Copernicus} & \citet{Branduardi1978}\\
46160.840(3)  & {\em EXOSAT} & \citet{Haberl1989}\\
48722.94(31)  & {\em Granat} & \citet{Rubin1996}\\
48651.365(31) & {\em BATSE} & \citet{Rubin1996}\\
49149.425(27)  & {\em BATSE} & \citet{Rubin1996}\\
50783.650(15) & \rxte/ASM & Present Work \\
52175.579(15) & \rxte/ASM & Present Work \\
52861.29(2)  & {\em INTEGRAL} & Present work\\
53270.69(2)  & {\em INTEGRAL} & Present work\\
53574.341(15)& \rxte/ASM & Present Work \\
53472.00(2) & {\em INTEGRAL} & Present work \\
53785.82(3)  & {\em INTEGRAL} & Present work\\
54164.55(2)  & {\em INTEGRAL} & Present work\\
54341.97(3)  & {\em INTEGRAL} & Present work\\
54962.83(2) & \rxte/ASM & Present Work \\
\hline
\end{tabular}
\end{centering}
\end{table}

\begin{table}[h]
\tiny
\begin{centering}
\begin{tabular}{lll}
\hline 
\multicolumn{3}{c}{4U 1538-522}\\ 
\hline 
Epoch Time&  Satellite & Reference\\
$T_{\pi/2}$ &&\\
\hline
\hline 
43015.8(1)  & {\em OSO-8} & \citet{Becker1977}\\
45517.660(50) & {\em Tenma} & \citet{Makishima1987} \\
47221.474(20)  & {\em Ginga} & \citet{Corbet1993}\\
48600.979(27)  & {\em BATSE} & \citet{Rubin1997}\\
49003.629(22) & {\em BATSE} & \citet{Rubin1997}\\
49398.855(29)   & {\em BATSE} & \citet{Rubin1997}\\
49797.781(22)  & {\em BATSE} & \citet{Rubin1997}\\
50450.206(14)  & {\em RXTE} & \citet{Clark2000} \\
52851.33(1)  & {\em RXTE} & \citet{Mukherjee2006}\\
52855.042(25)  & {\em RXTE} & \citet{Baykal2006}\\
\hline
\hline
$T_{ecl}$ &  &\\ 
\hline
\hline 
41449.95(7)  & {\em Uhuru} & \citet{Cominsky1991}\\
42628.0(1) & {\em Ariel-V}  & \citet{Davison1977}\\
43258.35(10)  & {\em Ariel-V} & \citet{Davison1977}\\
43384.889(22)  & {\em HEAO-A1} & \citet{Cominsky1991}\\
43563.827(22)  & {\em HEAO-A1} & \citet{Cominsky1991}\\
45920.2(2) & {\em EXOSAT} & \citet{Cominsky1991}\\
45923.9(2) &   {\em EXOSAT} & \citet{Cominsky1991}\\
51016.96(3) & \rxte/ASM & Present Work \\
52702.22(3)  & {\em INTEGRAL}& Present Work \\
52877.43(3) & \rxte/ASM & Present Work \\
53779.695(50)  & {\em INTEGRAL} & Present Work\\
54734.18(3) & \rxte/ASM & Present Work \\
54868.37(5)  & {\em INTEGRAL} & Present Work\\
\hline
\end{tabular}
\end{centering}
\end{table}

\begin{table}[h]
\tiny
\begin{centering}
\begin{tabular}{lll}
\hline
\multicolumn{3}{c}{SMC X-1}\\ 
\hline
Epoch Time&  Satellite & Reference\\
$T_{\pi/2}$ &&\\
\hline 
\hline
40963.99(2)  & {\em Uhuru} & \citet{Schreier1972}\\
42275.65(4)  & {\em Copernicus} &  \citet{Tuohy1975}\\
42836.1828(2)  & {\em SAS-3} &  \citet{Primini1977}\\
42999.6567(16)  & {\em Ariel 5} &  \citet{Davison1977}\\
43116.4448(22) & {\em COS-B} &  \citet{BonnetBidaud1981} \\
46942.47237(15)  & {\em Ginga} &  \citet{Levine1993}\\
47401.744476(7)  & {\em Ginga} & \citet{Levine1993}\\
47740.35906(3)  & {\em Ginga} & \citet{Levine1993}\\
48534.34786(35)  & {\em ROSAT} &  \citet{Wojdowski1998}\\
48892.4191(5)  & {\em ROSAT} &  \citet{Wojdowski1998}\\
49102.59109(82)  & {\em ASCA} & \citet{Wojdowski1998}\\
49137.61911(5) & {\em ROSAT} &  \citet{Wojdowski1998}\\
50091.170(63)  & {\em RXTE} &  \citet{Wojdowski1998}\\
50324.691861(8)  & {\em RXTE} &  \citet{Inam2010}\\
50787.849(14) & \rxte/ASM & Present Work \\
51694.67302(1)  & {\em RXTE} & \citet{Raichur2010}\\
52185.052(3) & \rxte/ASM & Present Work \\
52846.700(25)  & {\em INTEGRAL} & Present Work\\
52979.017(1)  & {\em RXTE} & \citet{Raichur2010}\\
53582.268(14) & \rxte/ASM & Present Work \\
\hline
\end{tabular}
\end{centering}
\end{table}

\begin{table}[h]
\tiny
\begin{centering}
\begin{tabular}{lll}
\hline
\multicolumn{3}{c}{SAX J1802.7-2017}\\ 
\hline
Epoch Time&  Satellite & Reference\\
$T_{\pi/2}$ &&\\
\hline 
\hline 
52168.22(12) & {\em BeppoSAX} & \citet{Augello2003}\\
52168.26(4)  & {\em BeppoSAX} & \citet{Hill2005}\\
\hline
\hline
$T_{ecl}$ &  &\\ 
\hline
52931.37(2) & {\em INTEGRAL} &  \citet{Hill2005}\\
53260.37(7) & {\em INTEGRAL} & \citet{Jain2009J}\\
53776.82(7) & {\em Swift} & \citet{Jain2009J}\\
53863.1(14)  & {\em INTEGRAL} & Present Work\\
54503.38(7) & {\em Swift} & \citet{Jain2009J}\\
\hline
\end{tabular}
\end{centering}
\end{table}

\begin{table}[h]
\tiny
\begin{centering}
\begin{tabular}{lll}
\hline
\multicolumn{3}{c}{XTE J1855-026}\\ 
\hline
Epoch Time&  Satellite & Reference\\
$T_{\pi/2}$ &&\\
\hline 
51495.25(2)  & {\em RXTE} & \citet{Corbet2002}\\
\hline
$T_{ecl}$ &  &\\ 
\hline
52704.038(50)  & {\em INTEGRAL} & Present Work\\
54009.972(50)  & {\em INTEGRAL} & Present Work\\
54890.679(50)  & {\em INTEGRAL} & Present Work\\
\hline
\end{tabular}
\end{centering}
\end{table}

\begin{table}[h]
\tiny
\begin{centering}
\begin{tabular}{lll}
\hline
\multicolumn{3}{c}{Vela X-1}\\ 
\hline
Epoch Time&  Satellite & Reference\\
$T_{\pi/2}$ &&\\
\hline 
\hline
42611.23(6)  & {\em SAS-3} & \citet{Rappaport1976}\\
42727.750(24)  & {\em COS-B} & \citet{vanderKlis1984}\\
42996.628(19) & {\em COS-B} & \citet{vanderKlis1984}\\
43821.34(13) & {\em SAS-3} & \citet{Rappaport1980} \\
44170.937(21) & {\em COS-B} & \citet{vanderKlis1984}\\
44278.5466(36)&{\em ECHELEC}& \citet{vanKerkwijk1995a}\\
44305.44(7) & {\em Hakucho} & \citet{Nagase1981}\\
45408.056(13)  & {\em Hakucho} & \citet{Nagase1983}\\
48895.2186(12)  & {\em BATSE} & \citet{Bildsten1997}\\
52974.008(8)  & {\em INTEGRAL} & \citet{Kreykenbaum2008}\\
\hline
\hline
$T_{ecl}$ &  &\\ 
\hline
41446.04(7) & {\em Uhuru} & \citet{Forman1973} \\
42449.97(2) &  {\em Copernicus} &  \citet{Charles1978}\\
42620.32(3) &    {\em Ariel-V} & \citet{Watson1977}\\
42727.93(3) &  {\em COS-B} & \citet{Ogelman1977}\\
43651.11(7) &    {\em OSO-8} & \citet{Deeter1987b}\\
44314.57(5) &   {\em Hakucho} & \citet{Nagase1983}\\
50777.947(16) & {\em RXTE/ASM} &Present work \\
52176.395(16)  & {\em RXTE/ASM} &  Present work\\
52803.90(2)  & {\em INTEGRAL} & Present work\\
52812.88(2) & {\em INTEGRAL} &  Present work \\
52974.225(3) & {\em INTEGRAL} &Present work \\
52983.17(3)  & {\em INTEGRAL} &  Present work\\
53574.834(16) & {\em RXTE/ASM} &Present work \\
54964.333(16)  & {\em RXTE/ASM} &  Present work\\
\hline
\end{tabular}
\end{centering}
\end{table}

\begin{table}[h]
\tiny
\begin{centering}
\begin{tabular}{lll}
\hline
\multicolumn{3}{c}{EXO 1722-363}\\ 
\hline
Epoch Time&  Satellite & Reference\\
$T_{\pi/2}$ &&\\
\hline 
\hline
51112.187(22) & {\em RXTE} & \citet{Thompson2007}$^{*}$\\
51219.350(6) & {\em RXTE} & \citet{Corbet2005}\\
52875.276(92) & {\em RXTE} & \citet{Thompson2007}$^{*}$\\
53761.679(24) & {\em RXTE} & \citet{Thompson2007}$^{*}$\\
53976.083(97)  & {\em XMM-Newton} & \citet{Manousakis2011}$^{*}$\\
\hline
$T_{ecl}$ &  &\\ 
\hline
51219.35(5) & {\em RXTE} & \citet{Corbet2005}\\
51988.91(4) & {\em RXTE}  & \citet{Markwardt2003}\\
52670.8(1) & {\em INTEGRAL} & \citet{ZuritaHeras2006}\\
53079.885(50)  & {\em INTEGRAL} &  Present work\\
53761.69(1)  & {\em INTEGRAL} & \citet{Manousakis2011}\\
53985.715(50)  & {\em INTEGRAL} &  Present work\\
\hline 
\end{tabular}
\end{centering}
\end{table}

\begin{table}[h]
\tiny
\begin{centering}
\begin{tabular}{lll}
\hline
\multicolumn{3}{c}{OAO 1657-415}\\ 
\hline 
Epoch&  Satellite & Reference\\
$T_{\pi/2}$ &&\\
\hline
\hline 
48390.6549(27) & {\em BATSE} & \citet{Jenke2011}\\
48515.99(5)  & {\em BATSE} & \citet{Chakrabarty1993}\\
48547.3800(52) & {\em BATSE} & \citet{Jenke2011}\\
48578.7293(50) & {\em BATSE} & \citet{Jenke2011}\\
48735.4386(33) & {\em BATSE} & \citet{Jenke2011}\\
49299.5984(62) & {\em BATSE} & \citet{Jenke2011}\\
49623.4633(48) & {\em BATSE} & \citet{Jenke2011}\\
50260.7701(26) & {\em BATSE} & \citet{Jenke2011}\\
50292.1121(52) & {\em BATSE} & \citet{Jenke2011}\\
50323.4677(31) & {\em BATSE} & \citet{Jenke2011}\\
50354.8047(21) & {\em BATSE} & \citet{Jenke2011}\\
50584.6562(45) & {\em BATSE} & \citet{Jenke2011}\\
50689.116(50)  & {\em RXTE} & \citet{Baykal2011}\\
52663.893(10)  & {\em INTEGRAL} & \citet{Barnstedt2008}\\
54721.7666(30) & {\em Fermi} & \citet{Jenke2011}\\
54753.1092(34) & {\em Fermi} & \citet{Jenke2011}\\
55254.5552(25) & {\em Fermi} & \citet{Jenke2011}\\
55306.7930(25) & {\em Fermi} & \citet{Jenke2011}\\
55338.1411(31) & {\em Fermi} & \citet{Jenke2011}\\
55526.1813(24) & {\em Fermi} & \citet{Jenke2011}\\
55776.9135(38) & {\em Fermi} & \citet{Jenke2011}\\
\hline
$T_{ecl}$ &  &\\ 
\hline
52674.158(60)  & {\em INTEGRAL} & Present work\\
52705.4(1)  & {\em INTEGRAL} & \citet{Denis2005}\\
52883.07(10) & \rxte/ASM & Present Work \\
53186.04(10)  & {\em INTEGRAL} & Present work\\
53562.173(60)  & {\em INTEGRAL} & Present work\\
\hline
\end{tabular}
\end{centering}
\end{table}

\end{appendix}  

\clearpage
\bibliographystyle{aa}
\bibliography{references}

\end{document}